\IfFileExists{svjour3.cls}{%
  \documentclass[twocolumn,epjc3]{svjour3}
  \smartqed
  \newif\ifepjcclass\epjcclasstrue
}{%
  \documentclass[twocolumn,10pt]{article}
  \usepackage[margin=0.72in]{geometry}
  \newif\ifepjcclass\epjcclassfalse
}

\usepackage[utf8]{inputenc}
\usepackage[T1]{fontenc}
\usepackage{lmodern}
\usepackage{graphicx}
\usepackage{amsmath,amssymb,amsfonts,mathtools}
\usepackage{bm}
\usepackage{booktabs}
\usepackage{array}
\usepackage{multirow}
\usepackage{enumitem}
\usepackage{caption}
\usepackage{subcaption}
\usepackage{microtype}
\usepackage{xcolor}
\usepackage{hyperref}
\usepackage{cite}
\usepackage{placeins}
\usepackage{float}

\graphicspath{{figures/}}
\hypersetup{colorlinks=true,linkcolor=blue,citecolor=blue,urlcolor=blue}

\newcommand{\dd}{\mathrm{d}}
\newcommand{\xb}{x_B}
\newcommand{\ReH}{\operatorname{Re}\mathcal{H}}
\newcommand{\ReE}{\operatorname{Re}\mathcal{E}}
\newcommand{\ReHt}{\operatorname{Re}\widetilde{\mathcal{H}}}
\newcommand{\sigDVCS}{\sigma_{\mathrm{DVCS}}}
\newcommand{\CFF}{\mathcal{F}}

\newcommand{\surr}{\mathrm{surr}}
\newcommand{\alg}{\mathrm{alg}}
\newcommand{\meth}{\mathrm{meth}}
\newcommand{\expd}{\mathrm{exp}}
\newcommand{\tot}{\mathrm{tot}}
\newcommand{\truth}{\mathrm{true}}
\newcommand{\rep}{\mathrm{rep}}
\newcommand{\uu}{\mathrm{UU}}
\newcommand{\lu}{\mathrm{LU}}
\newcommand{\cov}{\mathrm{cov}}

\ifepjcclass
\journalname{Eur. Phys. J. C}
\title{Experimental uncertainty propagation for neural-network extraction of Compton form factors}
\author{D. Keller\thanksref{e1}}
\thankstext{e1}{e-mail: dustin@virginia.edu}
\institute{Department of Physics, University of Virginia, Charlottesville, Virginia 22904, USA}
\else
\title{Experimental uncertainty propagation for neural-network extraction of Compton form factors}
\author{D. Keller\\\small Department of Physics, University of Virginia, Charlottesville, Virginia 22904, USA\\\small \texttt{dustin@virginia.edu}}
\date{}
\fi

\begin{document}
\maketitle

\begin{abstract}
Neural-network extractions of deeply virtual Compton scattering (DVCS) Compton form factors (CFFs) require uncertainty propagation through the full inference chain.  We present a CFF-specific framework for four complementary workflows: local extraction from binned angular data, kinematic multivariate inference (KMI), symmetry-constrained observable surrogates, and event-level or re-binnable analyses.  A deterministic reduced Belitsky--Kirchner--Mueller (BKM) layer maps network outputs to observables, while covariance-weighted residuals define the loss.  Experimental covariance, stochastic training variability, methodological variation, binning and support effects, and signed closure bias are retained at matrix or kernel level.  Nested replica--training ensembles directly separate experimental and algorithmic covariance, while a complementary non-nested construction combines one fit per replica with fixed-data retraining for stability and non-Gaussianity studies.  The displayed four-parameter closure extraction uses unpolarized cross sections and outputs $(\ReH,\ReE,\ReHt,\sigDVCS)$.  Beam-spin-asymmetry surfaces are used only to demonstrate odd-$\phi$ surrogate symmetry; a joint cross-section/asymmetry extraction requires the relevant imaginary CFFs.  Hall A-like local pseudodata and surface-level pseudodata closure studies illustrate non-Gaussian replicas, conservative coverage, and weak four-CFF identifiability rather than providing phenomenological CFF constraints.  Direct surface-replica propagation is preferred, with mean-surface/kernel compression serving as a second-moment alternative.  We also describe detector covariance, re-binning, bin migration, and tiered experimental data products.  The resulting prescription specifies central values, covariance matrices or kernels, bias and coverage diagnostics, and the statistical interfaces needed for a versioned global surrogate ensemble that can be updated as new DVCS measurements become available.
\ifepjcclass
\else
\fi
\end{abstract}

\section{Scope and motivation}\label{sec:intro}

Generalized Parton Distributions (GPDs) encode correlations between longitudinal parton momentum and transverse spatial structure.  DVCS,
\begin{equation}
  e(k)+p(P)\rightarrow e(k')+p(P')+\gamma(q'),
\end{equation}
provides one of the cleanest experimental channels for accessing GPD information through CFFs, which are convolutions of GPDs with perturbatively calculable coefficient functions \cite{Mueller1994,Ji1997D,Ji1997PRL,Radyushkin1996,Radyushkin1997}.  The inverse problem is controlled by several coupled sources of uncertainty: finite experimental statistics, detector resolution, correlated normalization and calibration uncertainties, bin migration, background subtraction, incomplete covariance information, theory reductions in the observable layer, and neural-network training variability.  A neural network can represent CFFs or observables flexibly \cite{Cybenko1989,Hornik1991,Goodfellow2016}, but the central value has limited physics meaning unless these uncertainty sources are propagated to the final CFFs and reported in a form that can be reproduced.

The present paper develops a methodological guide for CFF extraction from DVCS observables, with the central objective of defining an uncertainty budget that can be applied consistently across local fits, KMI, surrogate-assisted extraction, and event-level or re-binnable analyses.  The use of neural networks together with Monte Carlo replicas has a well-developed precedent in PDF determinations, where the NNPDF program introduced neural-network representations of structure functions and parton distributions with replica-based uncertainty propagation and closure-test validation \cite{DelDebbio2005,DelDebbio2007,Ball2009,NNPDFClosure2014,BallNNPDF2022}.  Related neural-network analyses in TMD and hadronic inverse problems provide additional context for data-driven partonic inference \cite{Fernando2023,BacchettaMAP2025,Fernando2025TMD}, while the formalism and numerical studies below are built around the DVCS/CFF problem.  {The statistical tools used throughout comprise covariance replicas, informed-bootstrap ensembles, covariance-weighted losses, closure tests, and model-selection diagnostics.  They are organized into a CFF-specific workflow that defines the quantities to be computed, their propagation through the analysis, and their presentation.}

{The framework is organized around three practical objectives: preserving the correlations required for downstream uncertainty propagation, identifying bias and coverage failures through closure tests, and enabling independent reproduction of the reported uncertainty budget.  Because the level of experimental information available for release can vary substantially among analyses, Sec.~\ref{sec:datarelease} presents a tiered hierarchy of data products that supports these objectives at progressively increasing levels of detail.}

These objectives also have a cumulative purpose.  They provide the statistical and data-product prerequisites for a versioned global surrogate ensemble of DVCS observables, trained on the available world data and updated as new measurements become available.  In a community implementation, each release would preserve experiment-specific covariance, nuisance responses, measurement conditions, provenance, and data-support information rather than collapse the measurements into a single unqualified smooth curve.  Each validated version would provide an auditable snapshot of the current data-supported observable representation from which local or KMI CFF extractions could be regenerated consistently.  The present work does not construct this world-data surrogate; it establishes the uncertainty bookkeeping, validation criteria, and interoperable data products required for such an effort.

Four extraction regimes are treated.  First, local extraction fixes $x=(x_B,t,Q^2)$ and extracts a reduced CFF vector from the $\phi$ dependence of one or more observables.  Second, kinematic multivariate inference (KMI) trains a single CFF model over many kinematic settings with the same BKM observable loss, eliminating the need to stitch together independent local CFF points.  Third, an experimental-data surrogate learns symmetry-constrained smooth observable surfaces from sparse data replicas and passes a distribution of those surfaces to either local fits or KMI.  Fourth, when event-level or re-binnable information is available, detector-level covariance, recursive re-binning, and bin-migration studies are folded directly into the replica generation. The analysis therefore explicitly builds the uncertainty budget for all four regimes. This list of extraction methodologies is not exhaustive. We selected only the most basic types of extraction methods to illustrate the uncertainty discussion.  {The Hall A-like pseudodata closure study supplies the numerical examples in the figures and tables and is used to demonstrate where the uncertainty definitions succeed or fail; no physical CFF result is inferred from those values.}

The reduced BKM formalism used in the examples is the common observable layer for these studies.  It provides a practical map from CFFs to photon electroproduction observables and has been used in recent CFF studies, including global deep-neural-network modeling constrained from local maps and quantum deep-neural-network CFF extraction \cite{CaleroDiazKeller2025,LeKeller2026}.  Earlier neural-network CFF parameterizations \cite{Kumericki2011,Moutarde2019,Cuic2020} motivate the use of neural networks as flexible CFF representations; the emphasis here is the propagation, decomposition, and reporting of uncertainty across several extraction architectures.

\section{Reduced BKM CFF extraction}\label{sec:bkm}

The photon electroproduction amplitude is decomposed as
\begin{equation}
  |\mathcal{T}|^2=|\mathcal{T}_{\mathrm{BH}}|^2+|\mathcal{T}_{\mathrm{DVCS}}|^2+\mathcal{I},
  \label{eq:amp}
\end{equation}
where BH denotes Bethe--Heitler photon emission from the lepton line, DVCS denotes photon emission from the hadronic subprocess, and $\mathcal I$ is the interference term.  The kinematic variables are $Q^2=-q^2$, $\xb=Q^2/(2P\cdot q)$, $t=(P'-P)^2$, lepton energy loss $y=(P\cdot q)/(P\cdot k)$, and the azimuthal angle $\phi$ between the leptonic and hadronic planes, defined with the Trento convention \cite{BacchettaTrento2004}.  For an unpolarized beam and target, the four-fold cross section is written schematically as
\begin{equation}
 \frac{\dd^4\sigma}{\dd \xb\,\dd Q^2\,\dd |t|\,\dd\phi}
 =\frac{\alpha^3\xb y^2}{8\pi Q^4\sqrt{1+\epsilon^2}}\frac{1}{e^6}|\mathcal T|^2,
 \label{eq:xsdef}
\end{equation}
with $\epsilon=2\xb M/Q$.  The BH term is fully calculable from QED and electromagnetic form factors, here represented with Kelly's parametrization \cite{Kelly2004}.  In BKM notation the unpolarized BH term has a finite harmonic expansion,
\begin{equation}
  |\mathcal{T}_{\mathrm{BH}}|^2 =
  \frac{e^6}{\xb^2 y^2(1+\epsilon^2)^2 t\,\mathcal P_1(\phi)\mathcal P_2(\phi)}
  \sum_{n=0}^{2}c_n^{\mathrm{BH}}\cos(n\phi),
\end{equation}
while the interference term has the structure
\begin{equation}
  \mathcal I=
  \frac{e^6}{\xb y^3 t\,\mathcal P_1(\phi)\mathcal P_2(\phi)}
  \sum_{n=0}^{3}c_n^{\mathcal I}\cos(n\phi) .
\end{equation}
The explicit coefficients are those of the BKM formalism \cite{BelitskyKirchnerMueller2002,BelitskyMueller2010}.  {The factors $\mathcal P_1(\phi)$ and $\mathcal P_2(\phi)$ are the two dimensionless lepton-propagator factors in the BKM convention.  The phrase ``BKM observable layer'' used below means the deterministic, non-trainable implementation of these BKM cross-section or asymmetry formulas that maps network-produced CFFs to observables; it is not an additional trainable neural-network layer.}  The reduced examples in this paper keep the helicity-conserving twist-2 real CFFs in the interference term and treat the pure DVCS contribution as an azimuthally independent fitted term.  The fitted local parameter vector is therefore
\begin{equation}
\begin{aligned}
  \theta(x)&=\left(\ReH(x),\ReE(x),\ReHt(x),\sigDVCS(x)\right),\\
  x&=(\xb,t,Q^2).
\end{aligned}
  \label{eq:thetavec}
\end{equation}
The three real CFF components are dimensionless.  The fourth fitted quantity, $\sigDVCS$, is the azimuthally independent contribution added directly to the four-fold unpolarized cross section and therefore has units of nb/GeV$^4$.  In the implementation used here, the Bethe--Heitler and interference terms include the explicit conversion $1~\mathrm{GeV}^{-2}=0.389379~\mathrm{mb}$, while the common leptoproduction prefactor supplies the remaining $\mathrm{GeV}^{-4}$ dependence.  Consequently, the BKM observable, its residuals and uncertainty components, and $\sigDVCS$ are all expressed in nb/GeV$^4$.

Double-helicity-flip gluon CFFs and effective twist-3 helicity-changing terms are neglected in the closure-test layer.  This reduction is useful for methodology because it preserves the nonlinear inverse problem while allowing error sources to be isolated.

For a fixed kinematic setting with $N_\phi$ angular measurements, the covariance-weighted observable loss is
\begin{equation}
{
\begin{aligned}
  \chi^2(\theta) &={} \bm d^{\,T}
  \left(\mathbf C^{\mathrm{data}}\right)^{-1}\bm d,\\
  d_i &={} O_{\mathrm{BKM}}(\theta;x,\phi_i)
  -O_i^{\mathrm{data}} .
\end{aligned}}
  \label{eq:chi2}
\end{equation}
{Here $O$ may be a cross section, asymmetry, or a stacked vector of observables, and $\mathbf C^{\mathrm{data}}$ denotes the experimental covariance in that same observable ordering.  The displayed four-parameter closure examples use only unpolarized cross-section entries.  If asymmetries or a stacked observable vector are fitted, $\theta$ and the deterministic BKM layer must be expanded consistently to include the CFF components to which those observables are sensitive, including the relevant imaginary CFFs.}  If only diagonal uncertainties are available, Eq.~\eqref{eq:chi2} reduces to the weighted mean-squared error
\begin{equation}
  L(\theta)=\frac{1}{N_\phi}\sum_{i=1}^{N_\phi}
  \frac{\left[O_{\mathrm{BKM}}(\theta;x,\phi_i)-O_i^{\mathrm{data}}\right]^2}{\delta_i^2}.
  \label{eq:wmse}
\end{equation}
{The symbol $\delta_i$ is the quoted one-standard-deviation uncertainty of observable entry $i$ when no off-diagonal covariance is available.}
The local closure-test examples use $N_\phi=24$ angular cross-section points at one fixed kinematic setting.  The network is trained at the observable level: it outputs CFFs whose BKM-predicted observable agrees with the measured or pseudodata observable.

\section{Replica propagation and error definitions}\label{sec:error_defs}

\subsection{Experimental covariance replicas}

{Let $\mathbf F=(F_1,\ldots,F_N)^T$ denote the ordered vector of measured observable central values.}  The replica construction used here is the finite-dimensional version of Monte Carlo integration over an experimental probability measure.  Monte Carlo sampling was introduced as a general statistical method for estimating quantities that are difficult to integrate analytically and is now a standard route for propagating uncertainty through nonlinear calculations \cite{MetropolisUlam1949,Press1992}.  Bootstrap resampling, introduced by Efron as a data-driven estimate of sampling distributions, provides the complementary frequentist picture in which repeated pseudo-experiments are generated from the available data \cite{Efron1979,DavisonHinkley1997}.  Bootstrap-based ensembles also connect naturally to machine-learning variance reduction and uncertainty estimation, as in bootstrap aggregation and modern deep ensembles \cite{Breiman1996,Lakshminarayanan2017}.  In a covariance-informed experimental analysis one usually uses a parametric, covariance-sampled replica rather than an uninformed resampling of binned points.  {With observable-space covariance $\mathbf C^{\mathrm{data}}$,} a statistically consistent Gaussian replica is generated by
\begin{equation}
  {\mathbf F^{(r)}=\mathbf F+\mathbf L\bm z^{(r)},\qquad
  \mathbf C^{\mathrm{data}}=\mathbf L\mathbf L^T,\qquad
  z_i^{(r)}\sim\mathcal N(0,1),}
  \label{eq:replica}
\end{equation}
where {$\mathbf L$} is a Cholesky factor \cite{Press1992,GolubVanLoan1996}.  A useful schematic covariance decomposition is
\begin{equation}
{
\begin{aligned}
 [\mathbf C^{\mathrm{data}}]_{ij}={}&
 \delta_{ij}\sigma_{\mathrm{stat},i}^2
 +\delta_{ij}\sigma_{\mathrm{abs},i}^2\\
 &+f_{\mathrm{sys},i}f_{\mathrm{sys},j}
 F_i^{\mathrm{ref}}F_j^{\mathrm{ref}}\\
 &+\sum_{\lambda}s_i^{[\lambda]}s_j^{[\lambda]} .
\end{aligned}}
 \label{eq:covmodel}
\end{equation}
{Here $\delta_{ij}$ is the Kronecker delta, $\sigma_{\mathrm{stat},i}$ is an uncorrelated statistical standard deviation, $f_{\mathrm{sys},i}$ is the fractional response to a multiplicative systematic, $F_i^{\mathrm{ref}}$ is a fixed reference prediction or nuisance-parameter value used to avoid scaling the variance with a fluctuated measurement, $\sigma_{\mathrm{abs},i}$ is an uncorrelated absolute systematic standard deviation, and $s_i^{[\lambda]}$ is the signed one-standard-deviation shift from correlated nuisance source $\lambda$.  Multiplicative normalizations may equivalently be fitted as nuisance parameters.}  In real applications the matrix should include all known statistical, normalization, calibration, acceptance, background, beam, target, reconstruction, and bin-migration contributions.  If the full covariance is unavailable, the final uncertainty budget must explicitly state which correlations were omitted and whether the resulting errors are likely underestimates.

{To avoid repeating a scalar variance estimator at every stage, for any replica output vector $\mathbf Z^{(r)}$ with component indices $\alpha,\beta$ and ensemble size $N_Z$ we use}
\begin{equation}
{
\begin{aligned}
 \overline Z_\alpha={}&\frac{1}{N_Z}
 \sum_{r=1}^{N_Z}Z_\alpha^{(r)},\\
 [\mathbf\Sigma_Z]_{\alpha\beta}={}&
 \frac{1}{N_Z-1}\sum_{r=1}^{N_Z}
 (Z_\alpha^{(r)}-\overline Z_\alpha)\\
 &\times(Z_\beta^{(r)}-\overline Z_\beta),\\
 s_{Z,\alpha}={}&\sqrt{[\mathbf\Sigma_Z]_{\alpha\alpha}}.
\end{aligned}}
 \label{eq:generic_sample_covariance}
\end{equation}
{Thus $\mathbf C^{\mathrm{data}}$ always denotes covariance in measured-observable space, $\mathbf\Sigma^{\mathrm{CFF}}$ denotes a finite-dimensional covariance among CFF components, and $K^O$ or $K^{\mathrm{CFF}}$ denotes a covariance kernel over observable or CFF surfaces.  Scalar widths quoted below are diagonal summaries of these full covariance objects, not replacements for them.}

Two operational implementations are common.  In an informed bootstrap, one trains an independent model on each $F^{(r)}$.  In a Monte Carlo replica fit, one trains a model with replica sampling embedded in the training or inference procedure.  If the same covariance information is used, the two approaches should agree up to algorithmic retraining variability.  The ensemble size $N_{\mathrm{rep}}$ is increased until the replica mean, variance, covariance, and final CFF intervals are stable.  The local closure study uses $N_{\mathrm{rep}}=1000$; the ensemble sizes used in the surface studies are stated explicitly in Secs.~\ref{sec:surrogate} and~\ref{sec:kmi}.

\subsection{Connection to neural-network PDF uncertainty propagation}\label{sec:replica_pdf_context}

The same statistical logic underlies the neural-network PDF methodology developed by NNPDF.  Early work first used neural networks and Monte Carlo replicas to construct unbiased representations of the proton structure function $F_2^p$ while retaining experimental errors and correlations \cite{DelDebbio2005}.  The method was then extended to parton distributions themselves, beginning with the nonsinglet case \cite{DelDebbio2007}, and subsequently to full PDF determinations in which a Monte Carlo ensemble of neural-network PDFs represented the probability measure induced by the experimental data \cite{Ball2009}.  In that setting each replica dataset is fitted independently, and the distribution of fitted neural-network functions supplies central values, variances, correlations, and derived observable uncertainties.  Closure testing then checks whether the fitting machinery can recover a known generating PDF within the assigned uncertainties \cite{NNPDFClosure2014,BallNNPDF2022}.

The present CFF workflow follows the same probability-measure viewpoint but changes the object being fitted.  Instead of fitting PDFs directly to inclusive observables, the neural network outputs CFFs, a BKM observable layer maps those CFFs to DVCS cross sections or asymmetries, and the covariance-weighted observable residual is minimized.  For local fits the replica ensemble produces a distribution of CFF values at fixed $x=(x_B,t,Q^2)$.  For KMI and surrogate-assisted KMI the ensemble produces correlated CFF surfaces, so the final object is not only a set of pointwise intervals but also a covariance kernel over kinematic phase space.  This distinction is important for later propagation to GPD-level quantities, benchmark comparisons, and theory constraints.

\subsection{Accuracy, precision, algorithmic error, and methodological error}

The uncertainty budget separates four quantities.  This separation is essential because a narrow DNN ensemble can be precise but biased, while a broad ensemble can have good coverage but poor resolving power.

{For CFF components $\CFF_a$ at fixed $x$, let $\widehat{\CFF}^{(r,q)}_a(x)$ be training repeat $q$ on experimental replica $r$, and let $\CFF_{a,\truth}(x)$ be the known truth in a closure test.  In the preferred nested design, the experimental covariance is computed from the algorithmically averaged result for each replica,}
\begin{equation}
{
\begin{aligned}
 \widetilde{\CFF}^{(r)}_a(x)={}&\frac{1}{N_{\alg}}
 \sum_{q=1}^{N_{\alg}}\widehat{\CFF}^{(r,q)}_a(x),\\
 \overline{\CFF}_a(x)={}&\frac{1}{N_{\mathrm{rep}}}
 \sum_{r=1}^{N_{\mathrm{rep}}}\widetilde{\CFF}^{(r)}_a(x),\\
 [\mathbf\Sigma^{\mathrm{CFF}}_{\expd}]_{ab}={}&
 \frac{1}{N_{\mathrm{rep}}-1}\sum_{r=1}^{N_{\mathrm{rep}}}
 \delta_r\CFF_a\,\delta_r\CFF_b,\\
 \delta_r\CFF_a={}&\widetilde{\CFF}^{(r)}_a-\overline{\CFF}_a,\\
 s_{a,\expd}={}&\sqrt{[\mathbf\Sigma^{\mathrm{CFF}}_{\expd}]_{aa}}.
\end{aligned}}
  \label{eq:cffprecision}
\end{equation}
The accuracy, or closure-test bias, is
\begin{equation}
  {b_a(x)=\overline{\CFF}_a(x)-\CFF_{a,\truth}(x),
  \qquad \epsilon_a(x)=|b_a(x)| .}
  \label{eq:bias}
\end{equation}
{The signed bias vector $\bm b$ is the primary closure diagnostic; $\epsilon_a$ is retained only as an optional non-negative scalar summary for tables.  An automatic bias correction is not applied unless the signed bias is stable under independent pseudodata generators and is itself calibrated with coverage tests.  Correlations of the bias among CFF components are retained through the sample covariance of the signed bias vectors across closure trials.}

Algorithmic error is the variability of the same extraction when the input data are held fixed.  Here this means training-path variability induced by independently varied weight initialization, minibatch ordering, stochastic optimizer updates, and stopping history while the architecture and the complete training prescription are held fixed; it is not merely floating-point nondeterminism.  In the nested ensemble it is the average within-replica covariance,
\begin{equation}
{
\begin{aligned}
 [\mathbf\Sigma^{\mathrm{CFF}}_{\alg}]_{ab}={}&
 \frac{1}{N_{\mathrm{rep}}}\sum_{r=1}^{N_{\mathrm{rep}}}
 \frac{1}{N_{\alg}-1}\sum_{q=1}^{N_{\alg}}
 \delta_q\CFF^{(r,q)}_a\,\delta_q\CFF^{(r,q)}_b,\\
 \delta_q\CFF^{(r,q)}_a={}&
 \widehat{\CFF}^{(r,q)}_a-\widetilde{\CFF}^{(r)}_a,\\
 s_{a,\alg}={}&
 \sqrt{[\mathbf\Sigma^{\mathrm{CFF}}_{\alg}]_{aa}}.
\end{aligned}}
  \label{eq:algerr}
\end{equation}
For a dedicated fixed-data diagnostic, the same equation is used with $N_{\mathrm{rep}}=1$.

Two complementary ensemble designs are used for different purposes.  The nested construction of Eqs.~\eqref{eq:cffprecision} and~\eqref{eq:algerr} is the preferred estimator because it measures the between-replica experimental covariance and the within-replica algorithmic covariance directly.  For targeted stability and non-Gaussianity studies, a computationally lighter non-nested diagnostic trains one stochastic model on each experimental replica and combines that covariance with a separate fixed-data retraining ensemble.  Denoting the covariance of the single-fit-per-replica outputs by $\mathbf\Sigma^{\mathrm{CFF}}_{\rep,\mathrm{comb}}$, the diagnostic approximation is
\begin{equation}
\begin{aligned}
 \mathbf\Sigma^{\mathrm{CFF}}_{\rep,\mathrm{comb}}
 &\simeq \mathbf\Sigma^{\mathrm{CFF}}_{\expd}
 +\mathbf\Sigma^{\mathrm{CFF}}_{\alg},\\
 \mathbf\Sigma^{\mathrm{CFF}}_{\expd,\mathrm{decomp}}
 &\simeq \mathbf\Sigma^{\mathrm{CFF}}_{\rep,\mathrm{comb}}
 -\mathbf\Sigma^{\mathrm{CFF}}_{\alg}.
\end{aligned}
\label{eq:non_nested_decomp}
\end{equation}
This diagnostic is used only when the single-fit-per-replica and fixed-data retraining ensembles employ the same architecture, optimizer configuration, stopping rule, initialization distribution, and numerical protocol, and when the algorithmic covariance is stable across representative experimental replicas.  The covariance difference must be positive semidefinite within sampling uncertainty; material negative modes or strong replica dependence invalidate the approximation and require the nested design.  This single-fit-per-replica construction is useful for large-ensemble non-Gaussianity, failed-training, and rapid stability studies, but it does not by itself separate experimental and algorithmic variability.

The algorithmic term should be small compared with the propagated experimental spread.  In the studies motivating the present workflow, a useful empirical working criterion has been that $s_{a,\alg}$ should contribute no more than about 10--20\% of the total standard deviation for a well-constrained component.  This number is not a universal statistical bound; it is a practical stability threshold based on repeated CFF closure tests.  Its purpose is consistent with the general bias--variance and ensemble-uncertainty literature: sensitivity to initialization, training order, and the optimization path should not dominate the uncertainty induced by the experimental probability measure \cite{Geman1992,Breiman1996,Lakshminarayanan2017}.  If this threshold is exceeded, the architecture, optimizer, stopping rule, or relative scaling of the data-fidelity and regularization terms should be retuned before assigning physics meaning to the result.  Averaging a validated deep ensemble is also a standard way to reduce this component; the residual ensemble spread must still be measured.

Throughout this manuscript, \emph{algorithmic variability} refers to stochastic training paths under a fixed model and fixed training prescription.  Changing the architecture depth or width, activation, optimizer family or prescribed hyperparameters, learning-rate schedule, regularization, validation strategy, stopping criterion, or symmetry implementation is instead treated as a methodological variation.

Methodological variants are likewise averaged over their algorithmic repeats before their between-method covariance is formed.  In a non-nested diagnostic, either $\mathbf\Sigma^{\mathrm{CFF}}_{\rep,\mathrm{comb}}$ is carried as one combined term or it is replaced by the compatible decomposition $\mathbf\Sigma^{\mathrm{CFF}}_{\expd,\mathrm{decomp}}+\mathbf\Sigma^{\mathrm{CFF}}_{\alg}$; the combined replica covariance and a separate algorithmic covariance are never added simultaneously.

Methodological error quantifies sensitivity to reasonable analysis choices: generator family, BKM reduction, architecture depth and width, activation, optimizer family or prescribed hyperparameters, learning-rate schedule, regularization form or strength, validation and stopping strategy, symmetry implementation, binning strategy, fit window, treatment of correlations, and model-selection criteria.  If $N_m$ variations are tested and each produces a {signed closure-test bias vector} $b_a^{(m)}(x)$, then
\begin{equation}
{
\begin{aligned}
 [\mathbf\Sigma^{\mathrm{CFF}}_{\meth}]_{ab}={}&
 \frac{1}{N_m-1}\sum_{m=1}^{N_m}
 [b_a^{(m)}-\overline b_a]\\
 &\times[b_b^{(m)}-\overline b_b],\\
 s_{a,\meth}={}&
 \sqrt{[\mathbf\Sigma^{\mathrm{CFF}}_{\meth}]_{aa}}.
\end{aligned}}
  \label{eq:metherr}
\end{equation}
For the local fits, the CFF generator parameters are varied and the extraction procedure is repeated to construct the residual distribution. The same method is applied to the surrogate surfaces below, except that the residual is now a function defined over phase space.  {The full sample covariance is retained; no outer product of separately estimated scalar standard deviations is substituted for it.}

\subsection{Additional error channels used in the final budget}

The four quantities above form the core diagnostic set, but a complete CFF analysis usually needs several additional terms.  These terms are best propagated as nuisance replicas or methodological variations whenever possible, because several of them can be correlated.

{\textit{Covariance-model uncertainty} enters when the observable covariance itself depends on nuisance choices, missing correlations, or incomplete separation of systematic sources.  If $\mathbf C^{\mathrm{data}}=\mathbf C^{\mathrm{data}}(\eta)$ depends on covariance parameters $\eta$, then plausible values $\eta_q$ are sampled or profiled and the extraction is repeated.  The induced CFF covariance is}
\begin{equation}
{
\begin{aligned}
 D_{a,q}(x)={}&\overline{\CFF}_{a}^{(\eta_q)}(x)-
 \langle\overline{\CFF}_{a}^{(\eta)}(x)\rangle_\eta,\\
 [\mathbf\Sigma^{\mathrm{CFF}}_{\cov}(x)]_{ab}={}&
 \frac{1}{N_\eta-1}\sum_{q=1}^{N_\eta}
 D_{a,q}(x)D_{b,q}(x),\\
 s_{a,\cov}(x)={}&
 \sqrt{[\mathbf\Sigma^{\mathrm{CFF}}_{\cov}(x)]_{aa}}.
\end{aligned}}
  \label{eq:cov_unc}
\end{equation}
{The admissible alternatives are specified in observable space and propagated through the BKM fit; Eq.~\eqref{eq:cov_unc} is therefore an induced CFF covariance, not an arbitrary extra covariance postulated directly in CFF space.  When the uncertainty of variance parameters is itself represented by a likelihood, an errors-on-errors construction provides a more formal alternative.\footnote{For a likelihood-based treatment of uncertain error parameters, see Ref.~\cite{Cowan2019}.}  This term is especially important when only diagonal publication errors are available.}

\textit{Normalization and correlated systematic uncertainty} can be handled either by including the appropriate outer products in {$\mathbf C^{\mathrm{data}}$} or by adding nuisance parameters $\nu_j$ to the observable prediction,
\begin{equation}
  {L(\theta,\nu)=\bm d(\theta,\nu)^T [\mathbf C_{\rm stat}^{\mathrm{data}}]^{-1}\bm d(\theta,\nu)}
  +\sum_j \left(\frac{\nu_j}{\sigma_{\nu_j}}\right)^2 .
  \label{eq:nuisance_loss}
\end{equation}
The nuisance-parameter ensemble should be propagated to the final CFF covariance rather than used only as a goodness-of-fit correction.

\textit{Support and extrapolation uncertainty} is relevant for KMI and surrogate fits.  Let $\rho(u)$ be a local density or distance-to-training-support measure in observable space.  Regions with poor support should either be masked from the reported CFF surface or assigned an additional term $s_{\rm supp}(u)$ determined from closure tests that remove or thin nearby data.  A surface that is accurate on a dense validation grid but biased near sparse boundaries should be quoted with this support term or masked outside the supported domain.

\textit{Model-selection and regularization uncertainty} comes from architecture size, activation choice, optimizer, stopping rule, regularization strength, imposed smoothness, and symmetry constraints.  These choices are included in $s_{\meth}$ when they are varied as part of the closure-test family.  Information criteria provide useful diagnostics for balancing fit quality against model complexity.  For a model with likelihood $\mathcal L$, $k$ effective parameters, and $N$ data points, the Akaike information criterion (AIC), Bayesian information criterion (BIC), and minimum-description-length (MDL) penalty may be written schematically as
\begin{equation}
  \mathrm{AIC}=-2\ln\mathcal L+2k,\qquad
  \mathrm{BIC}=-2\ln\mathcal L+k\ln N,
  \label{eq:aicbic}
\end{equation}
\begin{equation}
  \mathrm{MDL}\simeq -\ln\mathcal L+\frac{k}{2}\ln N,
  \label{eq:mdl}
\end{equation}
up to convention-dependent constants and coding choices \cite{Akaike1974,Schwarz1978,Rissanen1978,BurnhamAnderson2002,Bishop2006,Hastie2009}.  In neural-network CFF extraction these criteria should be interpreted as relative diagnostics within a specified model family because the effective number of parameters can be smaller than the raw number of weights after regularization, early stopping, and physics constraints.  If a single model is selected by AIC, BIC, MDL, validation loss, or cross-validation, the spread among near-optimal choices should still be reported as a methodological component.

\textit{Non-Gaussian and outlier uncertainty} is diagnosed from the replica ensemble itself.  When the ensemble is skewed, multimodal, or contaminated by failed trainings, Gaussian widths are insufficient.  The analysis should report robust intervals from percentiles, median absolute deviations, or trimmed ensembles, together with a failure fraction and the rule used to remove pathological replicas.

{The final CFF covariance is assembled at matrix level,}
\begin{equation}
{
\begin{aligned}
 \mathbf\Sigma^{\mathrm{CFF}}_{\tot}={}&
 \mathbf\Sigma^{\mathrm{CFF}}_{\expd}+
 \mathbf\Sigma^{\mathrm{CFF}}_{\alg}+
 \mathbf\Sigma^{\mathrm{CFF}}_{\meth}\\
 &+\mathbf\Sigma^{\mathrm{CFF}}_{\cov}+
 \mathbf\Sigma^{\mathrm{CFF}}_{\mathrm{bin}}+
 \mathbf\Sigma^{\mathrm{CFF}}_{\mathrm{supp}}\\
 &+\mathbf\Sigma^{\mathrm{CFF}}_{\mathrm{cross}},\\
 s_{a,\tot}={}&
 \sqrt{[\mathbf\Sigma^{\mathrm{CFF}}_{\tot}]_{aa}}.
\end{aligned}}
  \label{eq:total_nobias}
\end{equation}
{The cross term is set to zero only after paired or nested ensembles show that the relevant component shifts have negligible empirical cross-covariance.  Otherwise the measured cross-covariances are retained.  For the non-nested diagnostic of Eq.~\eqref{eq:non_nested_decomp}, either the combined replica covariance is retained as one term, or it is replaced by the compatible experimental decomposition plus the algorithmic covariance.  The combined replica covariance and a separate algorithmic covariance are never included together.}  A closure-test root-mean-square error including bias is
\begin{equation}
  {e_{a,\mathrm{closure}}(x)=\sqrt{s_{a,\tot}^2(x)+b_a^2(x)} .}
  \label{eq:rmse}
\end{equation}
For real data, the final interval should report the variance components and the estimated {signed} closure-test bias separately, rather than hiding bias inside a single number.

For the non-nested scalar summary below, the histogram width from one fit per replica is explicitly labeled as a combined experimental-plus-algorithmic spread.  Its diagnostic decomposition is
\begin{equation}
 s_{a,\expd,\mathrm{decomp}}=
 \sqrt{\max\!\left[
 \left(s_{a,\rep,\mathrm{comb}}^{\mathrm{hist}}\right)^2
 -s_{a,\alg}^2,0\right]}.
 \label{eq:scalar_non_nested_decomp}
\end{equation}
The scalar totals use the independence approximation because this summary does not contain the component cross-covariances; the preferred full covariance treatment uses the matrix- or kernel-level nested construction.

\begin{table*}[t]
\centering
\caption{Diagnostic scalar uncertainty summary for the non-nested local pseudodata closure study.  The Gaussian-fit and histogram widths, $s_{\rep,\mathrm{comb}}^{\mathrm{Gaus}}$ and $s_{\rep,\mathrm{comb}}^{\mathrm{hist}}$, each contain experimental and algorithmic variability because one stochastic model was trained per covariance replica.  The fixed-data retraining width $s_{\alg}$ is used in Eq.~\eqref{eq:scalar_non_nested_decomp} to obtain the diagnostic estimate $s_{\expd,\mathrm{decomp}}$.  The last two columns use the scalar independence approximation $s_{\tot}^2=s_{\expd,\mathrm{decomp}}^2+s_{\alg}^2+s_{\meth}^2$ and $e_{\mathrm{closure}}^2=s_{\tot}^2+|b|^2$.  No cross-covariances are inferred from this scalar summary; a full analysis uses nested covariance matrices or kernels and reports the signed bias vector.  The three real-CFF rows are dimensionless, while the $\sigDVCS$ row is in nb/GeV$^4$.}
\label{tab:local_budget_expanded}
\scriptsize
\setlength{\tabcolsep}{3.0pt}
\begin{tabular}{lcccccccc}
\toprule
Fitted output & $s_{\alg}$ & \shortstack{$s_{\rep,\mathrm{comb}}$\\(Gaus.)} & \shortstack{$s_{\rep,\mathrm{comb}}$\\(hist.)} & \shortstack{$s_{\expd,\mathrm{decomp}}$\\(hist.)} & $s_{\meth}$ & $|b|$ & $s_{\tot}$ & $e_{\mathrm{closure}}$ \\
\midrule
$\ReH$      & 0.030  & 0.320 & 0.645 & 0.644 & 0.730 & 0.611 & 0.974  & 1.150 \\
$\ReE$      & 0.020  & 0.178 & 0.310 & 0.309 & 0.100 & 0.050 & 0.326  & 0.330 \\
$\ReHt$     & 0.100  & 0.144 & 0.143 & 0.102 & 0.100 & 0.010 & 0.174  & 0.175 \\
$\sigDVCS$  & 0.0001 & 0.050 & 0.057 & 0.057 & 0.005 & 0.002 & 0.0572 & 0.0573 \\
\bottomrule
\end{tabular}
\end{table*}

{Table~\ref{tab:local_budget_expanded} is a scalar summary of the local pseudodata closure study, not a physical CFF result.  It identifies the Gaussian-fit and histogram widths as combined non-nested spreads and gives the corresponding scalar totals.  The $\ReH$ extraction is dominated by methodological and broad replica variation, whereas the fitted $\sigDVCS$ term is much more stable.  The nested covariance treatment replaces this scalar decomposition by the signed bias vector and full CFF covariance.}

\subsection{Uncertainty components by extraction stage}\label{sec:stage_taxonomy}

The four error metrics above are common to all workflows, but their operational meaning changes with the stage of the extraction.  For this reason the budget is organized by where the uncertainty enters:
\begin{enumerate}[leftmargin=*]
\item \textit{Experimental observable uncertainty}: statistical fluctuations, normalization errors, correlated systematics, false asymmetries, beam and target polarization uncertainties where relevant, acceptance, background subtraction, and detector-resolution effects.  These enter through the experimental covariance or through event-level resampling.
\item \textit{Binning and re-binning uncertainty}: bin migration, finite bin-width corrections, alternative $\phi$ starting points, alternative $x_B$ and $t$ bin widths, and the choice of which events enter a kinematic chunk.  This term becomes measurable when event-level or re-binnable data are available.
\item \textit{Surrogate-stage uncertainty}: residual bias of the surrogate surface relative to known truth in closure tests, experimental replica covariance, stochastic training-path variability under a fixed surrogate architecture and fixed training protocol, and the covariance kernel connecting neighboring surface points.
\item \textit{CFF-inference algorithmic uncertainty}: variation of the extracted CFFs under retraining with identical inputs, including initialization, optimizer path, early stopping, batch ordering, and finite numerical precision.
\item \textit{Methodological uncertainty}: dependence on BKM reduction, generator family used in closure tests, architecture and activation, optimizer family or prescribed hyperparameters, learning-rate and validation strategy, stopping criterion, regularization, symmetry implementation, observable selection, covariance assumptions, support definition, and KMI/local workflow choices.
\item \textit{Coverage and calibration}: agreement between nominal confidence intervals and empirical closure-test coverage.  A quoted $68\%$ interval should cover the truth in approximately $68\%$ of statistically independent closure trials over the domain on which the method will be used.
\end{enumerate}
These components have distinct meanings.  For example, a surrogate residual surface is a bias diagnostic, while a replica precision surface is a repeated-experiment spread.  Algorithmic error from retraining and experimental covariance propagation therefore appear as separate entries in the budget.  The tables in this paper therefore keep the terms separate until the final reporting stage, where quadrature combinations are used only for components that have been tested for approximate independence.  {Table~\ref{tab:method_budget_map} maps each workflow to the covariance object and closure diagnostic needed to preserve this distinction.}

\begin{table*}[t]
\centering
\caption{Method-dependent uncertainty budget.  {Each row lists the quantities whose reporting is motivated by covariance preservation, bias diagnosis, or reproducibility for the corresponding extraction workflow.}}
\label{tab:method_budget_map}
\setlength{\tabcolsep}{3.2pt}
\begin{tabular}{p{0.16\textwidth}p{0.28\textwidth}p{0.24\textwidth}p{0.24\textwidth}}
\toprule
Workflow & Primary propagated object & Required closure diagnostics & Required reporting objects \\
\midrule
Local fit & CFF vector at fixed $(x_B,t,Q^2)$ & CFF bias, replica precision, algorithmic spread, generator-variation residuals & CFF mean, covariance matrix, $s_{\expd}$, $s_{\alg}$, $s_{\meth}$, bias, coverage \\
KMI & CFF surfaces over kinematics & Pointwise and domain coverage, smoothness-bias tests, regularization scans & CFF-surface mean, CFF covariance kernel, benchmark-point table, domain-averaged and maximum errors \\
Surrogate-assisted & Observable surfaces and then CFFs & Surface residual, surface precision, surface algorithmic error, surface methodological error & Observable-surface error table, covariance kernel, propagated CFF table \\
Event-level/re-binnable & Event replicas, bin statistics, CFFs & Detector-smearing closure, bin-migration closure, re-binning stability & Detector covariance assumptions, binning covariance $C_{\rm bin}$, CFF budget with re-binning term \\
\bottomrule
\end{tabular}
\end{table*}

\section{Local CFF extraction at fixed kinematics}\label{sec:local}

A local fit fixes $x=(\xb,t,Q^2)$ and extracts Eq.~\eqref{eq:thetavec} from the $\phi$ dependence of the observable.  The closure study uses the Jefferson Lab Hall A E00-110 unpolarized DVCS cross-section measurement of Defurne \textit{et al.} as the experimental scale for the pseudodata uncertainties \cite{Defurne2015}.  More recent Hall A measurements at high $x_B$ motivate multi-observable and multi-kinematic extensions \cite{Georges2022}.  In this local study the published binned information is used as the available data product, and the covariance is therefore treated as diagonal in $\phi$ unless an explicit correlation model is introduced.  This reproduces a common publication-level analysis condition and provides a common point of comparison for KMI, surrogate-assisted, and event-level analyses.

The pseudodata generator used for local closure tests is
\begin{equation}
  G(\xb,t)=\left(a\xb^2+b\xb\right)\exp\left(ct^2+dt+e\right)+f,
  \label{eq:gen}
\end{equation}
where $G$ denotes one of $\ReH,\ReE,\ReHt$, or $\sigDVCS$.  The parameters are chosen to produce distinguishable nonzero CFFs and are chosen for closure testing rather than as a physical GPD model.  {Table~\ref{tab:generator} lists the generator coefficients used for the displayed pseudodata examples.}

\begin{table}[t]
\centering
\caption{Parameters used in Eq.~\eqref{eq:gen} for the local pseudodata generator.  The momentum-transfer variable $t$ is expressed in GeV$^2$; consequently $c$ and $d$ have units GeV$^{-4}$ and GeV$^{-2}$, respectively, while $e$ is dimensionless.  The coefficients $a$, $b$, and $f$ carry the units of the generated output: dimensionless for the three real-CFF rows and nb/GeV$^4$ for the $\sigDVCS$ row.}
\label{tab:generator}
\setlength{\tabcolsep}{3pt}
\begin{tabular}{lrrrrrr}
\toprule
Output & $a$ & $b$ & $c$ & $d$ & $e$ & $f$ \\
\midrule
$\ReH$ & -4.41 & 1.68 & -9.14 & -3.57 & 1.54 & -1.37 \\
$\ReE$ & 144.56 & 149.99 & 0.32 & -1.09 & -148.49 & -0.31 \\
$\ReHt$ & -1.86 & 1.50 & -0.29 & -1.33 & 0.46 & -0.98 \\
$\sigDVCS$ & 0.50 & -0.41 & 0.05 & -0.25 & 0.55 & 0.166 \\
\bottomrule
\end{tabular}
\end{table}

\begin{figure}[t]
\centering
\includegraphics[width=\linewidth]{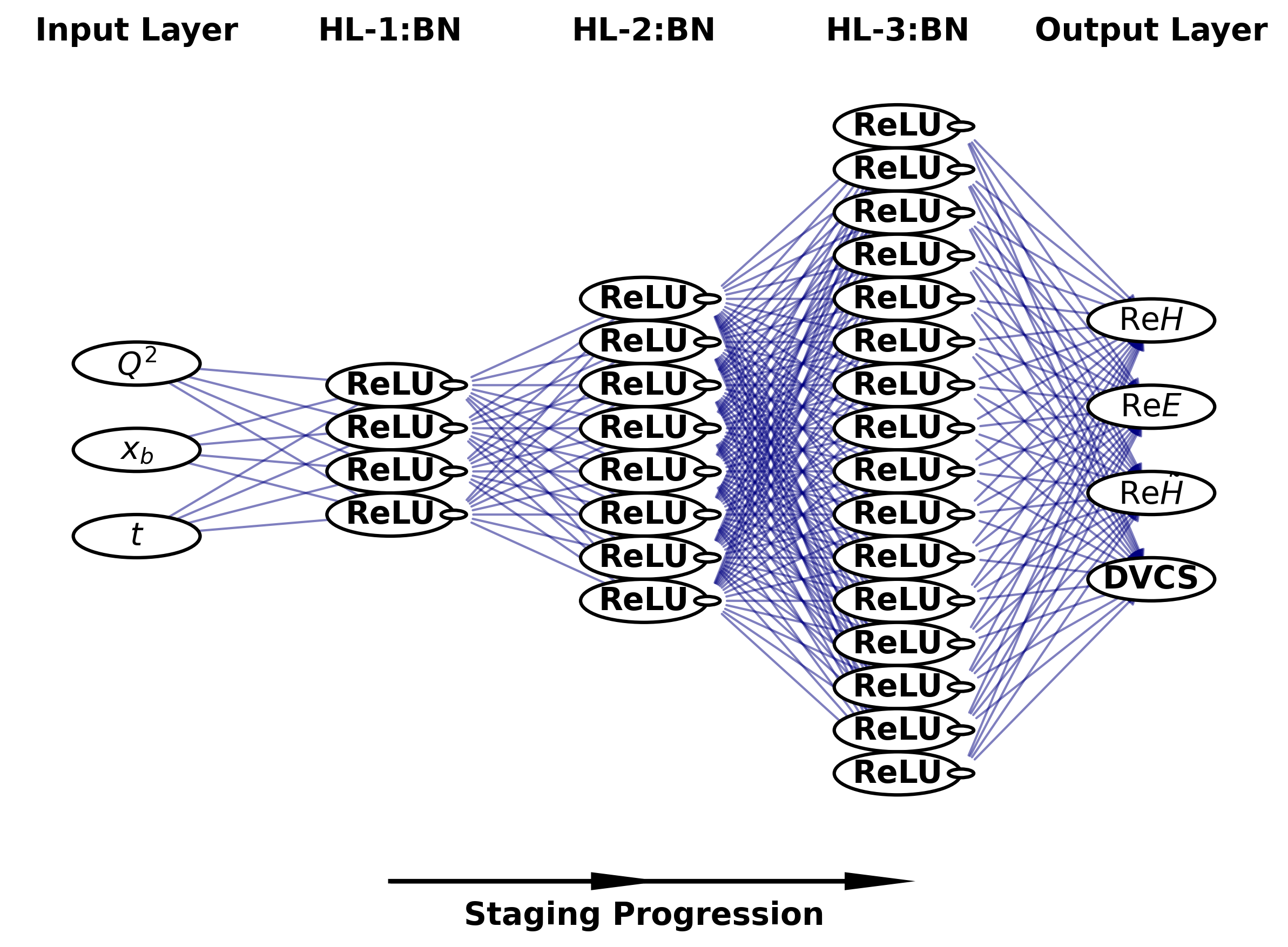}
\caption{Progressive DNN architecture used for the local closure study.  {The displayed widths $(4,8,16)$ are compressed for legibility; the implemented widths are 32, 64, 128, and 256 hidden nodes as specified in the text.}}
\label{fig:progressive}
\end{figure}

{The local neural network used in the closure study is a progressively expanded feed-forward model implemented with TensorFlow/Keras\footnote{\url{https://www.tensorflow.org/}}.  It has three inputs $(Q^2,\xb,t)$ and four linear outputs $(\ReH,\ReE,\ReHt,\sigDVCS)$.  The implemented hidden widths are 32, 64, 128, and 256; each dense hidden layer is followed by batch normalization and a ReLU activation.  The model is rebuilt from scratch at each stage rather than transferring weights from the preceding stage, and weights are initialized from a normal distribution with mean zero and standard deviation 0.1.  Each stage is trained for at most 100 epochs with Adam, an initial learning rate of 0.01 reduced by 10\% every 10 epochs, and early stopping with patience 10.  The batch size is one kinematic sample, whose target vector contains all 24 angular cross-section points, so the full angular distribution is evaluated simultaneously in the physics loss.  Independent random initializations are used in the retraining ensembles described in Sec.~3.3.  The deterministic BKM map converts the outputs into the 24 cross-section predictions entering Eq.~\eqref{eq:wmse}.  The physics restrictions are an unpolarized beam and target, helicity-conserving twist-2 real CFFs in the interference term, an azimuthally independent fitted DVCS term, no double-helicity-flip gluon CFFs, and no effective twist-3 helicity-changing terms.  Figure~\ref{fig:progressive} is a width-compressed schematic rather than a literal drawing of the full node counts.}

\begin{figure}[t]
\centering
\includegraphics[width=\linewidth]{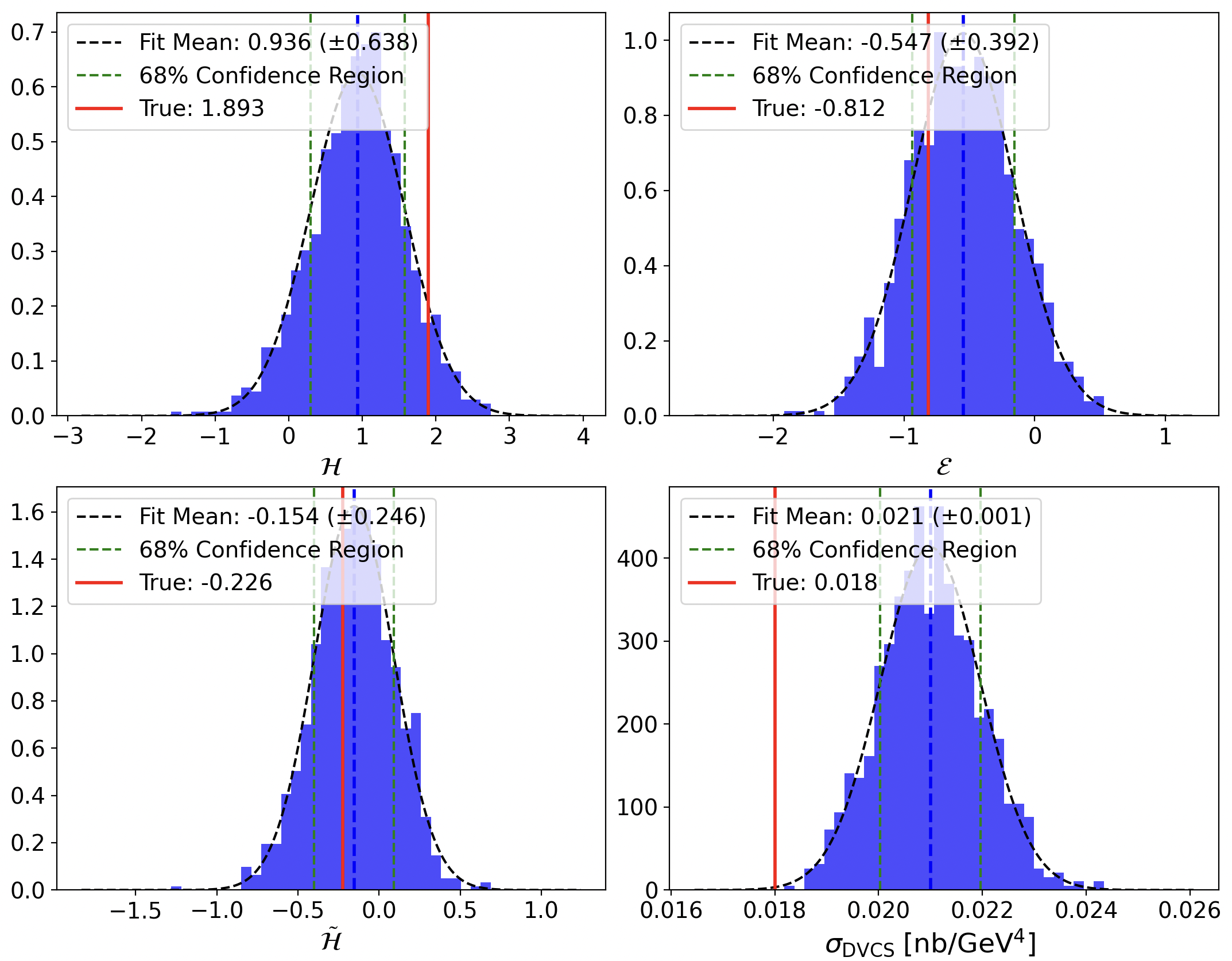}
\caption{Local CFF and DVCS-term extraction from a Hall A-like pseudodata closure test at $Q^2=2.091~\mathrm{GeV}^2$, $\xb=0.4$, $t=-0.371~\mathrm{GeV}^2$, and $E_{\mathrm{beam}}=5.75~\mathrm{GeV}$.  The red line is the generator truth and the blue dashed line is the fitted mean.  The three real-CFF panels are dimensionless, while the $\sigDVCS$ panel is in nb/GeV$^4$.  The figure illustrates why each fitted output requires its own accuracy and precision diagnostic.}
\label{fig:local1}
\end{figure}

\begin{figure}[t]
\centering
\includegraphics[width=\linewidth]{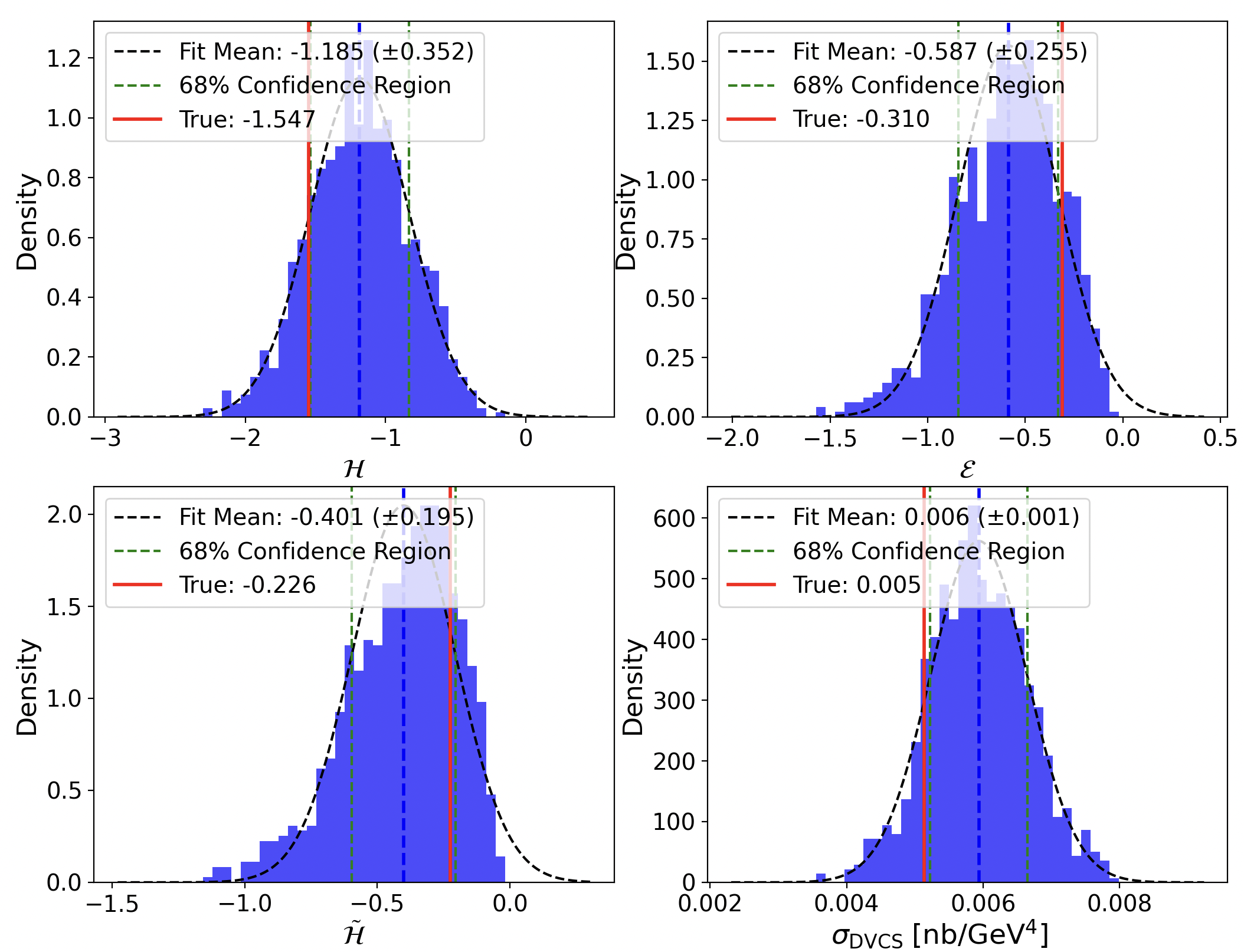}
\caption{Same local procedure and unit conventions as Fig.~\ref{fig:local1}, but with a different generator variation.  Repeating this procedure over many plausible generators produces the residual distribution used to estimate methodological error.}
\label{fig:local2}
\end{figure}

\begin{figure}[t]
\centering
\includegraphics[width=\linewidth]{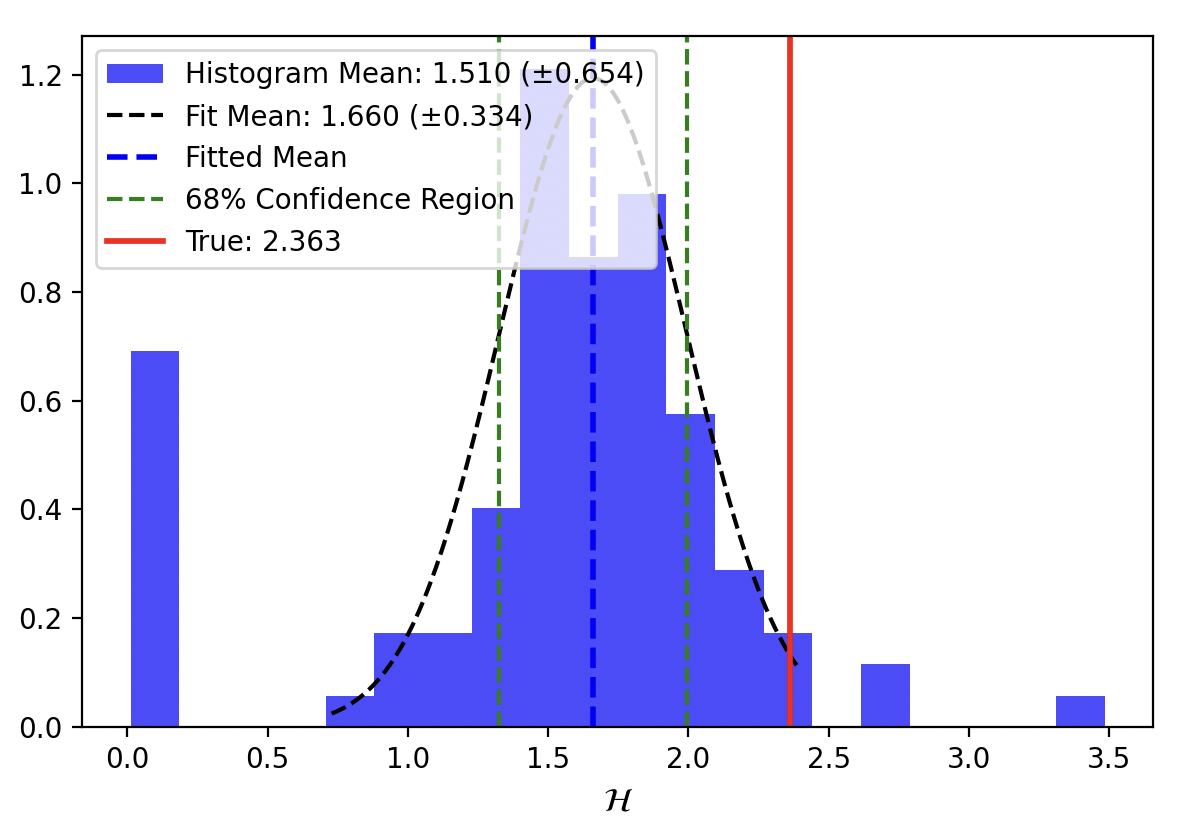}
\caption{Non-Gaussian $\ReH$ replica distribution in which failed or outlying trainings distort the histogram mean.  When replica distributions are non-Gaussian, percentile intervals, robust estimators, and Gaussian-fit widths should be quoted together.  Outlier diagnostics should also be reported.}
\label{fig:outlier}
\end{figure}

\begin{figure}[t]
\centering
\includegraphics[width=\linewidth]{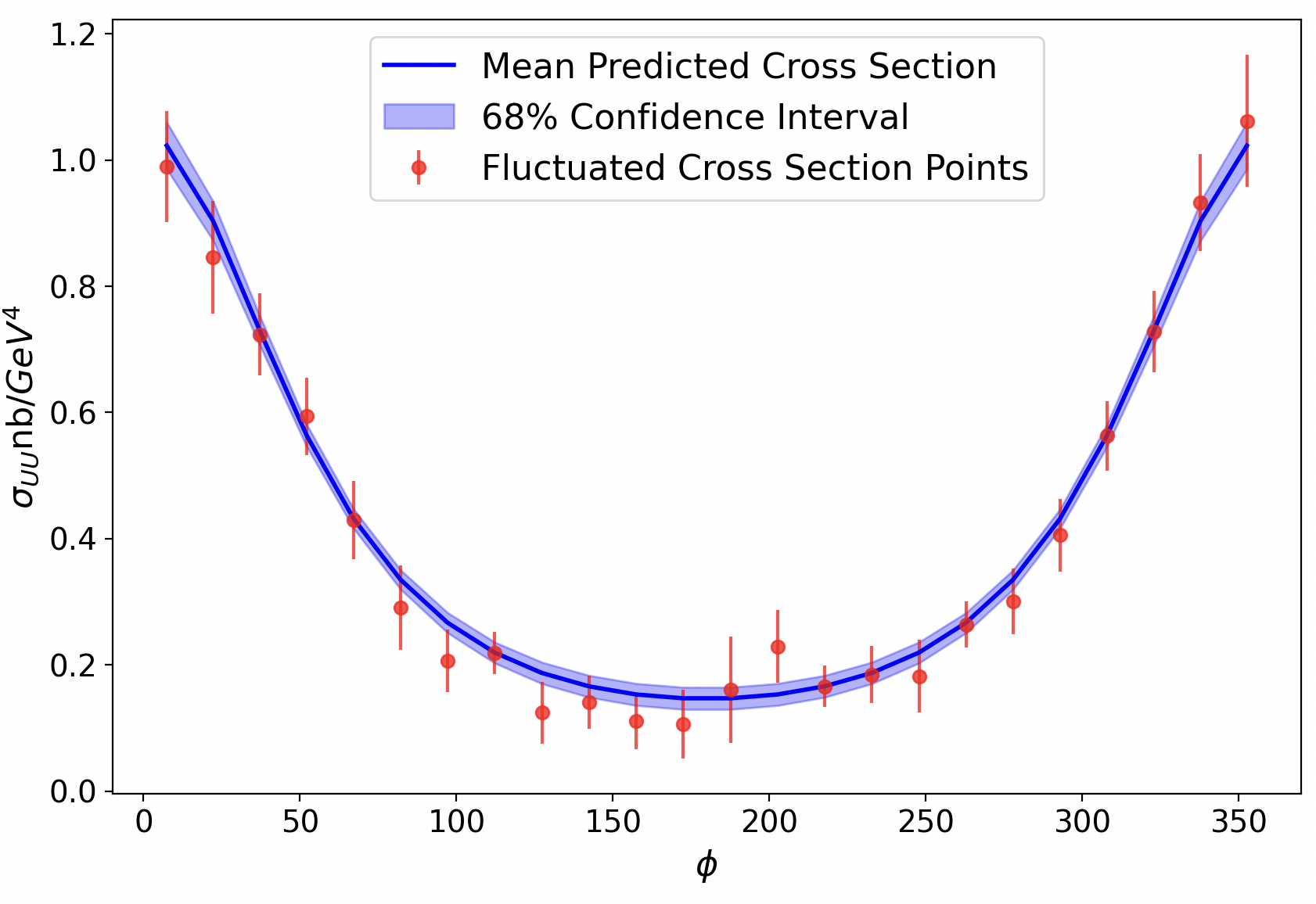}
\caption{Local cross-section reconstruction from a CFF replica ensemble.  This observable-level agreement is an important diagnostic, while different CFF combinations can still produce visually similar cross sections.  CFF-level closure tests and parameter-correlation matrices are required.}
\label{fig:crossx}
\end{figure}

{Figures~\ref{fig:local1} and~\ref{fig:local2} show two generator choices, Fig.~\ref{fig:outlier} demonstrates why a Gaussian width alone can be misleading, and Fig.~\ref{fig:crossx} shows that observable-level agreement does not by itself resolve CFF degeneracy.}  The local pseudodata study uses the non-nested variance-component construction of Eq.~\eqref{eq:non_nested_decomp}.  First, a pseudodata generator supplies the truth CFFs and BKM cross sections at the chosen kinematics.  Second, one unsmeared dataset is fitted repeatedly to measure the fixed-data algorithmic covariance.  Third, one stochastic model is trained on each covariance-sampled replica, producing the combined experimental-plus-algorithmic spread reported in Table~\ref{tab:local_budget_expanded}.  Pairing those two ensembles gives the diagnostic scalar decomposition of Eq.~\eqref{eq:scalar_non_nested_decomp} after the compatibility checks described in Sec.~\ref{sec:error_defs}.  This construction is useful for studying non-Gaussian tails, outliers, failed trainings, and numerical stability.  The preferred nested analysis trains repeated realizations within every experimental replica and uses the estimators of Eqs.~\eqref{eq:cffprecision} and~\eqref{eq:algerr}.  The generator parameters are then varied around a nominal value, the full extraction is repeated, and the signed residual distribution defines the methodological covariance in Eq.~\eqref{eq:metherr}.  The table uses 200 generator variations.  For a quantitative methodological assignment, the number of variations is increased until the residual moments and coverage estimates are stable.  Finally, the extraction is checked for non-Gaussian behavior and CFF correlations, with percentile intervals and robust summaries reported when Gaussian widths are inadequate.

For Hall A-like cross-section errors, relative uncertainties of order 15--35\% in the publication-level bins provide a useful scale for judging the resulting CFF precision.  This criterion should be applied in absolute rather than relative terms for CFFs near zero.  Observable-level agreement, such as the reconstructed cross section in Fig.~\ref{fig:crossx}, is an important diagnostic, but it does not resolve parameter degeneracy or CFF non-identifiability.  The local result should therefore report the CFF mean, precision, algorithmic error, methodological error, closure-test bias, CFF correlation matrix, and coverage for each CFF component.

\section{Experimental-data surrogate models}\label{sec:surrogate}

\subsection{Role of the surrogate in the uncertainty chain}

Experimental DVCS data are often sparse in $(\xb,t,Q^2,\phi)$, and the published information may not include full correlations.  A surrogate model provides a differentiable experimental representation of the measured observable,
\begin{equation}
  \widehat O_\psi(x,\phi),\qquad x=(\xb,t,Q^2),
\end{equation}
that can be evaluated on a dense grid, sampled by a later CFF extraction, and constrained by known symmetries.  In the uncertainty chain the surrogate sits between the experimental data product and the CFF-inference model.  Its task is to preserve the measured covariance while producing a smooth representation that can be used by local fits, KMI, or event-level re-binning studies.

A world-data implementation would enlarge the surrogate inputs and metadata beyond $x$ and $\phi$ to include every measurement condition required by the observable layer, such as beam energy or $y$, beam charge and helicity, target state, and observable definition.  Compatible measurements could share a physical representation, while experiment-specific normalization and systematic effects would remain encoded through covariance blocks and nuisance parameters rather than unconstrained experiment labels.  Updating the ensemble with a new dataset would require versioned retraining and revalidation of closure, coverage, support, and inter-dataset tension diagnostics for both the new and previously included measurements.  The global surrogate should therefore be understood as a maintained ensemble with provenance and uncertainty, not as a single fixed network or a replacement for the source data.

The surrogate stage has its own uncertainty budget.  Different covariance replicas of the experimental data lead to different surrogate surfaces; this is the propagated experimental component.  Repeated trainings of the same replica under a fixed architecture, optimizer configuration, learning-rate schedule, validation strategy, stopping rule, regularization, and symmetry implementation probe stochastic training-path variability; this is the algorithmic interpolation component.  Changes to any of those prescriptions, including network capacity or optimizer family and hyperparameters, are methodological variations.  Pseudodata surfaces generated from alternative plausible BKM/CFF truth models provide an additional methodological test.  The surrogate is accepted only on the data-supported domain where its closure-test residual is small compared with the experimental precision surface or is explicitly retained as a signed bias diagnostic.

\subsection{Symmetry-projected outputs}

The observable figures and the reduced BKM layer use the physical azimuthal angle
\(\phi\) on the interval \([0,2\pi]\), or equivalently
\(\phi_{\mathrm{deg}}\in[0,360^\circ]\) in the plots.  This convention keeps the
cross-section minimum near \(\phi_{\mathrm{deg}}=180^\circ\), as in the usual Hall
A-style angular distributions.  For symmetry arguments it is sometimes useful
to introduce the centered angle
\begin{equation}
  \varphi = \phi-\pi,\qquad \varphi\in[-\pi,\pi],
  \label{eq:centered_phi}
\end{equation}
so that \(\varphi=0\) corresponds to \(\phi=\pi\).  The paper uses this
centered variable only to state symmetry properties; the plotted observable
figures use the physical \(0\)--\(360^\circ\) angle unless explicitly stated
otherwise.

In the physical convention the reflection angle is
\begin{equation}
  \bar\phi=2\pi-\phi,
  \qquad
  \bar\phi_{\mathrm{deg}}=360^\circ-\phi_{\mathrm{deg}} .
  \label{eq:physical_phi_reflection}
\end{equation}
Equivalently, in the centered convention \(\bar\varphi=-\varphi\).  Given an
unconstrained network \(f_\psi(x,\phi)\), an even observable is represented by
\begin{equation}
  f_\psi^{(+)}(x,\phi)=\frac{1}{2}\left[f_\psi(x,\phi)+f_\psi(x,\bar\phi)\right],
  \label{eq:even}
\end{equation}
while an odd observable is represented by
\begin{equation}
  f_\psi^{(-)}(x,\phi)=\frac{1}{2}\left[f_\psi(x,\phi)-f_\psi(x,\bar\phi)\right].
  \label{eq:odd}
\end{equation}
Thus an unpolarized cross section is constrained as
\begin{equation}
  \sigma_{\uu}(x,\phi)=\sigma_{\uu}(x,2\pi-\phi),
  \label{eq:sigma_phi_symmetry}
\end{equation}
and a beam-spin asymmetry is constrained as
\begin{equation}
  A_{\lu}(x,\phi)=-A_{\lu}(x,2\pi-\phi).
  \label{eq:bsa_phi_symmetry}
\end{equation}
In the centered variable these same relations read
\(\sigma_{\uu}(x,\varphi)=\sigma_{\uu}(x,-\varphi)\) and
\(A_{\lu}(x,\varphi)=-A_{\lu}(x,-\varphi)\).  A multi-output surrogate may share
a hidden representation but use separate projected heads for cross sections and
asymmetries.  This removes the unphysical odd component of \(\sigma_{\uu}\) and
the unphysical even component of \(A_{\lu}\) by construction.  In an
implementation the network may take \(\phi\), a normalized degree variable, or
periodic inputs such as \((\cos\phi,\sin\phi)\); the projection is applied by
evaluating the same unconstrained network at the reflected angle in
Eq.~\eqref{eq:physical_phi_reflection}.

\begin{figure}[t]
\centering
\includegraphics[width=\linewidth]{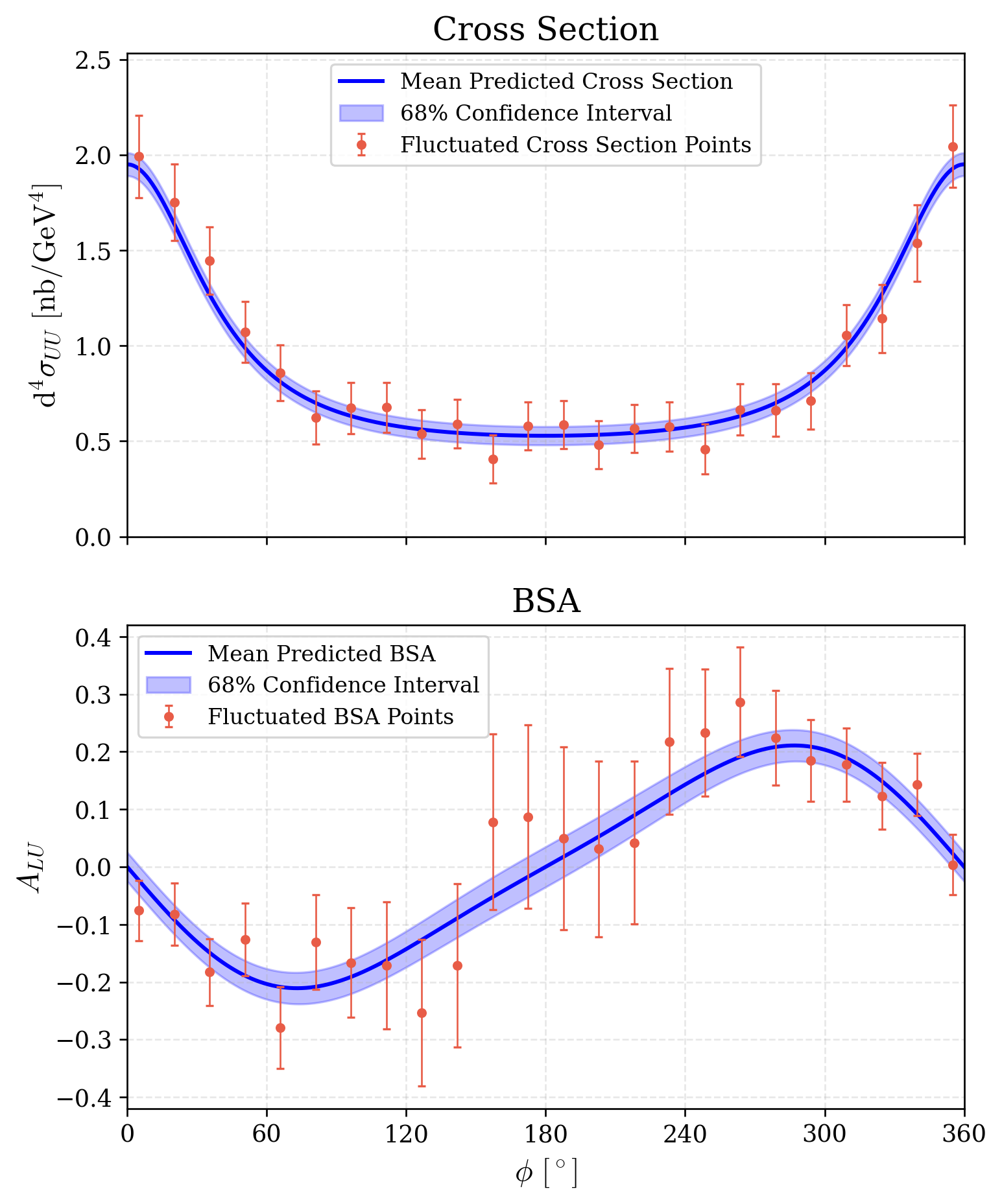}
\caption{Fixed-kinematics symmetry-constrained surrogate examples at $E_{\mathrm{beam}}=5.75~\mathrm{GeV}$, $Q^2=1.50~\mathrm{GeV}^2$, $x_B=0.22$, and $t=-0.20~\mathrm{GeV}^2$.  The upper panel shows the even unpolarized cross section in nb/GeV$^4$, calculated with the reduced BKM observable layer; the lower panel shows the dimensionless odd beam-spin asymmetry.  The larger BSA error bars reflect the uncertainty model assigned in this pseudodata example.}
\label{fig:surrogate_fixed_bkm}
\end{figure}

\begin{figure}[t]
\centering
\includegraphics[width=\linewidth]{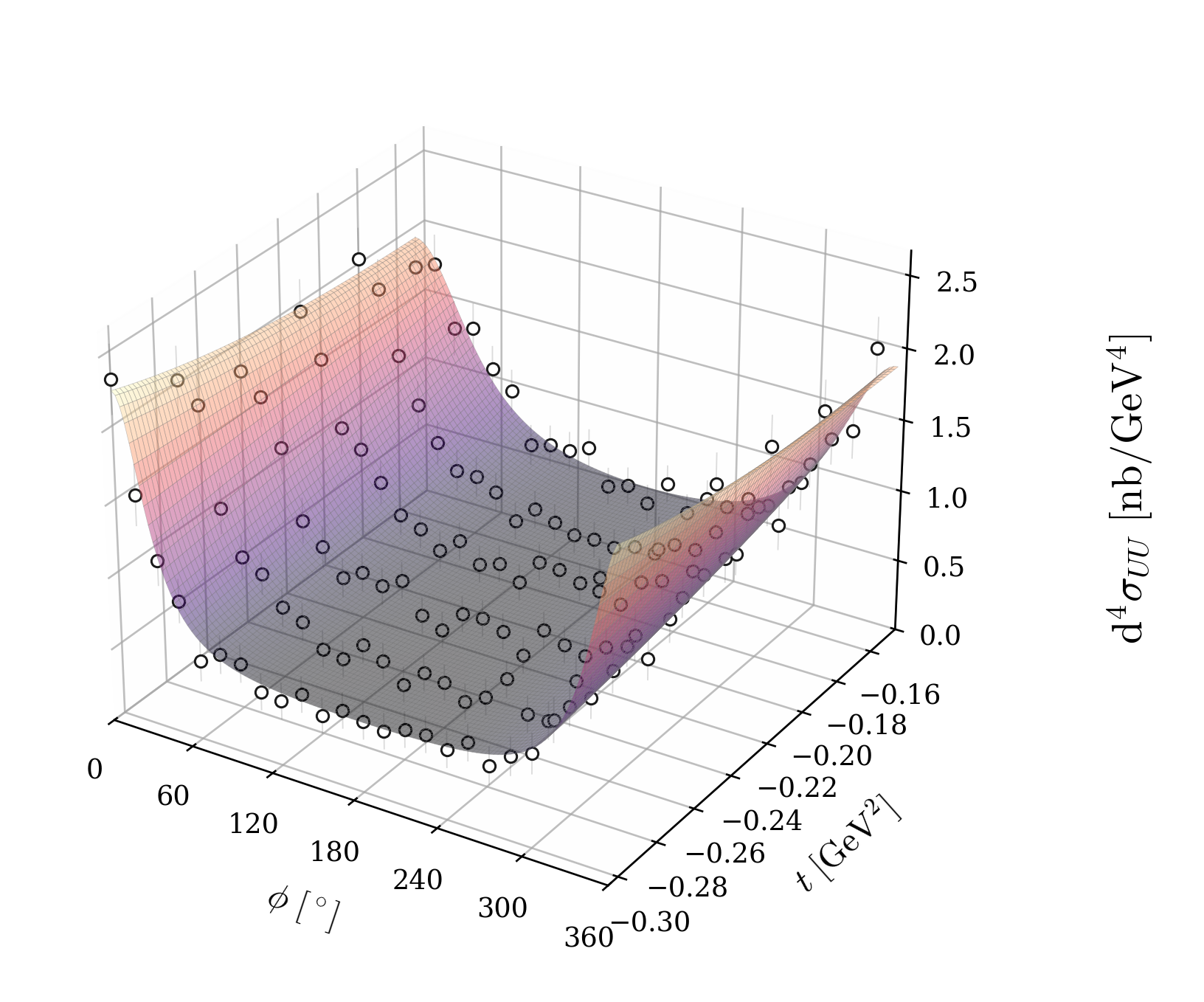}
\caption{Example of the unpolarized BKM cross-section surface as a function of $t$ and the physical azimuthal angle $\phi_{\mathrm{deg}}\in[0,360^\circ]$ at fixed $x_B=0.22$, $Q^2=1.50~\mathrm{GeV}^2$, and $E_{\mathrm{beam}}=5.75~\mathrm{GeV}$.  The vertical axis is $\dd^4\sigma_{UU}/(\dd x_B\,\dd Q^2\,\dd|t|\,\dd\phi)$ in nb/GeV$^4$.  The open circles are fluctuated pseudodata points at six discrete $t$ slices over $-0.30\leq t\leq-0.15~\mathrm{GeV}^2$.}
\label{fig:surrogate_surface_xs_bkm}
\end{figure}

\begin{figure}[t]
\centering
\includegraphics[width=\linewidth]{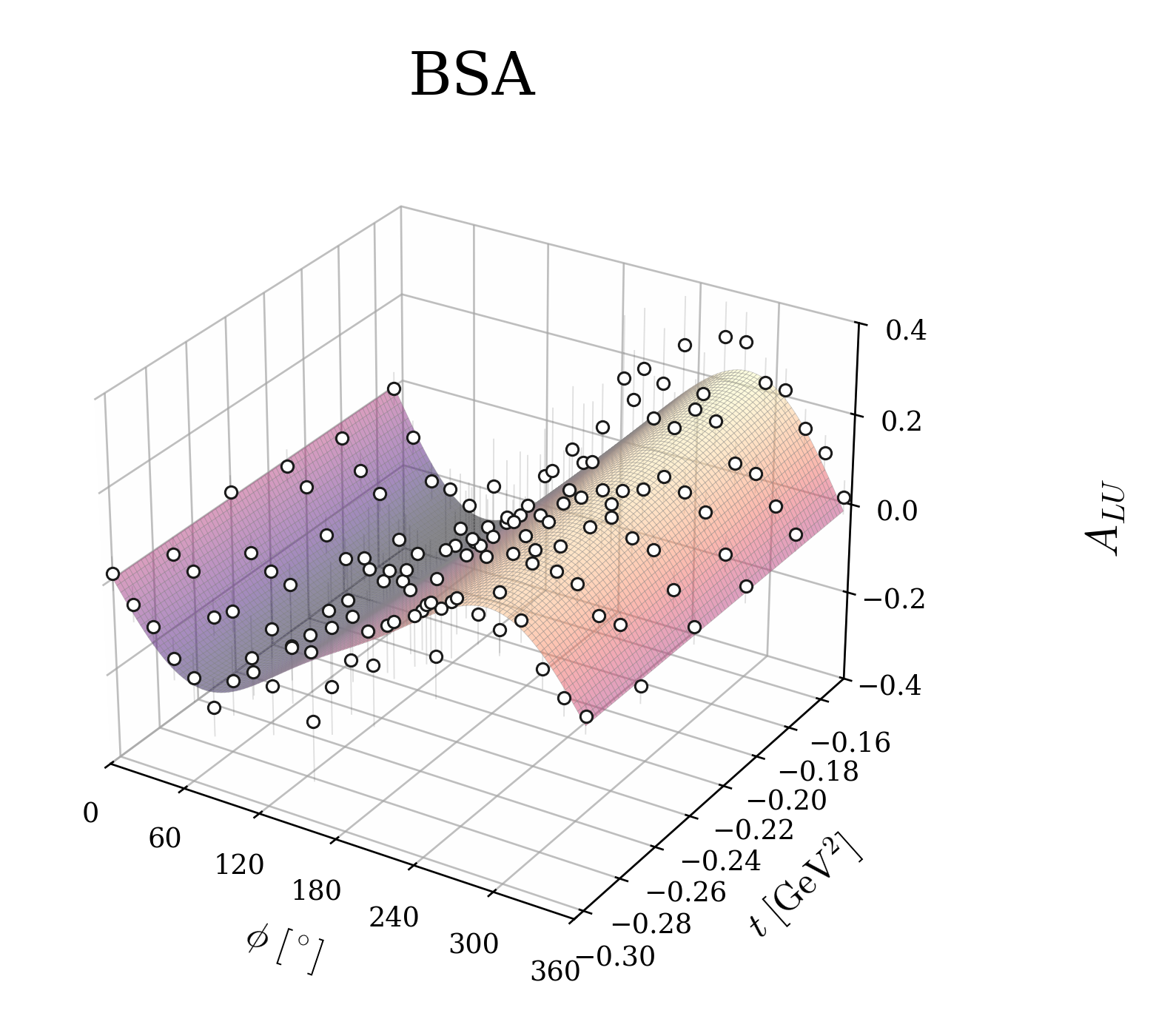}
\caption{Illustrative BSA surrogate surface in $(t,\phi_{\mathrm{deg}})$ at fixed $(x_B,Q^2)$ using the same physical-angle convention as Fig.~\ref{fig:surrogate_fixed_bkm}.  The vertical axis is the dimensionless beam-spin asymmetry $A_{LU}$.  The surface satisfies the odd reflection relation of Eq.~\eqref{eq:bsa_phi_symmetry}; open circles represent sparse pseudodata at discrete $t$ slices.}
\label{fig:surrogate_surface_bsa}
\end{figure}

{Figure~\ref{fig:surrogate_fixed_bkm} compares the even cross-section and odd beam-spin-asymmetry projections at fixed kinematics, while Figs.~\ref{fig:surrogate_surface_xs_bkm} and~\ref{fig:surrogate_surface_bsa} show how the same constraints extend over a sparse two-dimensional support.}

The BSA surfaces in Figs.~\ref{fig:surrogate_fixed_bkm} and~\ref{fig:surrogate_surface_bsa} are used only to demonstrate odd-$\phi$ symmetry and uncertainty propagation at the observable-surrogate stage.  They are not propagated through the displayed four-parameter CFF extraction of Eq.~\eqref{eq:thetavec}, which contains only real CFF components and an effective azimuthally independent DVCS term.  A joint extraction of $\sigma_{UU}$ and $A_{LU}$ in a full analysis would expand the CFF vector and the helicity-dependent BKM layer to include the relevant imaginary CFFs, construct the joint covariance across all observables, and use complementary beam-charge, beam-spin, or target-spin information where available.  Before additional components are freed, the weighted observable Jacobian or its singular-value spectrum should be examined to identify the combinations supported by the data; unsupported components must remain fixed or be controlled by explicitly documented priors or regularization.

\subsection{Surrogate closure test}

The surrogate should be configured with a closure test before it is used on real data.  The closure test uses a generator $O_{\truth}(u)$, with $u=(\xb,t,Q^2,\phi)$, to create sparse pseudodata at the same kinematic support and uncertainty scale as the experiment.  For the cross-section examples, $O_{\truth}$ is the reduced BKM observable evaluated with known CFF surfaces.  The closure-test procedure is:
\begin{enumerate}[leftmargin=*]
  \item Choose a truth surface $O_{\truth}(u)$ and an experimental design $\{u_i\}$ matching the real data support.
  \item Generate one pseudodata realization $O_i=O_{\truth}(u_i)+\eta_i$, with {$\eta\sim\mathcal N(0,\mathbf C^{\mathrm{data}})$}.
  \item Train a surrogate on the unsmeared or single-realization pseudodata and evaluate it on a dense validation grid $\mathcal G$.
  \item For the preferred nested design, generate $N_{\mathrm{rep}}$ covariance replicas, train $N_{\alg}$ stochastic surrogate realizations for every replica, and average within each replica before forming the experimental surface covariance.
  \item For the non-nested alternative used in stability and non-Gaussianity studies, train one surrogate per covariance replica and pair that ensemble with repeated trainings of a fixed dataset or representative replica subset.  Use the resulting decomposition only after the compatibility and positive-semidefiniteness checks of Eq.~\eqref{eq:non_nested_decomp}.
  \item Repeat the entire test for multiple truth surfaces, architectures, optimizer prescriptions, regularization strengths, validation strategies, and binning variants to estimate methodological error.
  \item Check coverage: the known truth should fall inside the quoted pointwise or simultaneous intervals with the stated frequency.
\end{enumerate}

{Let $\widehat O_{\psi,\alpha}^{(r,q)}(u)$ be observable component $\alpha$ of training repeat $q$ on experimental replica $r$.  The nested surface analogue of Eqs.~\eqref{eq:cffprecision} and~\eqref{eq:algerr} first averages over training repeats within each replica,}
\begin{equation}
{
\begin{aligned}
 \widetilde O_{\psi,\alpha}^{(r)}(u)={}&
 \frac{1}{N_{\alg}}\sum_{q=1}^{N_{\alg}}
 \widehat O_{\psi,\alpha}^{(r,q)}(u),\\
 \overline O_{\psi,\alpha}(u)={}&
 \frac{1}{N_{\mathrm{rep}}}\sum_{r=1}^{N_{\mathrm{rep}}}
 \widetilde O_{\psi,\alpha}^{(r)}(u),\\
 K^{O,\expd}_{\alpha\beta}(u,u')={}&
 \frac{1}{N_{\mathrm{rep}}-1}\sum_{r=1}^{N_{\mathrm{rep}}}
 \delta_r O_\alpha(u)\,\delta_r O_\beta(u'),\\
 \delta_r O_\alpha(u)={}&
 \widetilde O_{\psi,\alpha}^{(r)}(u)-\overline O_{\psi,\alpha}(u).
\end{aligned}}
\label{eq:surrogate_kernel}
\end{equation}
{The separately propagated algorithmic and methodological kernels are}
\begin{equation}
{
\begin{aligned}
 K^{O,\alg}_{\alpha\beta}(u,u')={}&
 \frac{1}{N_{\mathrm{rep}}}\sum_{r=1}^{N_{\mathrm{rep}}}
 \frac{1}{N_{\alg}-1}\\[-0.2em]
 &\quad\times\sum_{q=1}^{N_{\alg}}
 \delta_q O^{(r,q)}_\alpha(u)
 \delta_q O^{(r,q)}_\beta(u'),\\
 \delta_q O^{(r,q)}_\alpha(u)={}&
 \widehat O_{\psi,\alpha}^{(r,q)}(u)
 -\widetilde O_{\psi,\alpha}^{(r)}(u),\\
 K^{O,\meth}_{\alpha\beta}(u,u')={}&
 \frac{1}{N_m-1}\sum_{m=1}^{N_m}
 \delta_m R_{O,\alpha}^{(m)}(u)\\[-0.2em]
 &\quad\times\delta_m R_{O,\beta}^{(m)}(u'),\\
 \delta_m R_{O,\alpha}^{(m)}(u)={}&
 \widetilde R_{O,\alpha}^{(m)}(u)
 -\overline R_{O,\alpha}(u),\\
 s_{O,X,\alpha}(u)={}&
 \sqrt{K^{O,X}_{\alpha\alpha}(u,u)},\\
 X\in{}&\{\expd,\alg,\meth\}.
\end{aligned}}
\label{eq:surrogate_nested_kernels}
\end{equation}
{Here $\widetilde R_{O}^{(m)}$ is the algorithmically averaged residual surface for methodological variant $m$.  For a single observable the indices reduce to $\alpha=\beta=1$.  These definitions retain partial correlations among observable components and grid points without adding training-path variance twice or replacing the sample covariance by an outer product of scalar widths.}

The non-nested one-surrogate-per-replica alternative measures a combined surface kernel.  Under the same fixed-prescription, stability, and positive-semidefiniteness conditions as Eq.~\eqref{eq:non_nested_decomp},
\begin{equation}
\begin{aligned}
 K^{O,\rep,\mathrm{comb}}_{\alpha\beta}
 &\simeq K^{O,\expd}_{\alpha\beta}
 +K^{O,\alg}_{\alpha\beta},\\
 K^{O,\expd,\mathrm{decomp}}_{\alpha\beta}
 &\simeq K^{O,\rep,\mathrm{comb}}_{\alpha\beta}
 -K^{O,\alg}_{\alpha\beta}.
\end{aligned}
\label{eq:surrogate_non_nested}
\end{equation}
This approximation supports surface-stability and non-Gaussianity studies, while the nested kernel directly separates experimental and algorithmic covariance.

{The residual surface formed from the experimental-replica mean is}
\begin{equation}
  {R_{O,\alpha}(u)=\overline O_{\psi,\alpha}(u)-O_{\truth,\alpha}(u),}
  \label{eq:surrogate_residual}
\end{equation}
{Its absolute value is the pointwise surface accuracy in the closure test.  For a compact scalar summary of one observable component one may report}
\begin{align}
  {\mathrm{MAE}_{O,\alpha}=\frac{1}{|\mathcal G|}\sum_{u\in\mathcal G}|R_{O,\alpha}(u)|},\\
  {\mathrm{RMSE}_{O,\alpha}=\sqrt{\frac{1}{|\mathcal G|}\sum_{u\in\mathcal G}R_{O,\alpha}^2(u)}},\\
  {\mathrm{MaxE}_{O,\alpha}=\max_{u\in\mathcal G}|R_{O,\alpha}(u)|}.
\end{align}
{Here MAE is the mean absolute error, RMSE is the root-mean-square error, and MaxE is the maximum absolute error on the validation grid.  The scalar experimental, algorithmic, and methodological surfaces are the diagonal summaries $s_{O,\expd,\alpha}$, $s_{O,\alg,\alpha}$, and $s_{O,\meth,\alpha}$ of Eqs.~\eqref{eq:surrogate_kernel} and~\eqref{eq:surrogate_nested_kernels}.}
{The total closure-test surface error for one component is then}
\begin{equation}
{
\begin{aligned}
 e_{O,\mathrm{cl},\alpha}(u)={}&
 \Bigl[R_{O,\alpha}^2(u)+s_{O,\expd,\alpha}^2(u)\\
 &\qquad+s_{O,\alg,\alpha}^2(u)
 +s_{O,\meth,\alpha}^2(u)\Bigr]^{1/2}.
\end{aligned}}
  \label{eq:surrogate_total}
\end{equation}
{If one wants a quoted uncertainty band rather than an RMSE-like diagnostic, the bias term $R_{O,\alpha}^2$ is removed from the variance and reported separately.  Variance bands estimate repeated-experiment spread, while Eq.~\eqref{eq:surrogate_total} estimates closure-test error relative to known truth.  Ignoring the off-diagonal structure of $K^{O,X}_{\alpha\beta}(u,u')$ can double count information because neighboring grid evaluations and different observable outputs are correlated.}

\begin{figure*}[t]
\centering
\includegraphics[width=0.98\textwidth]{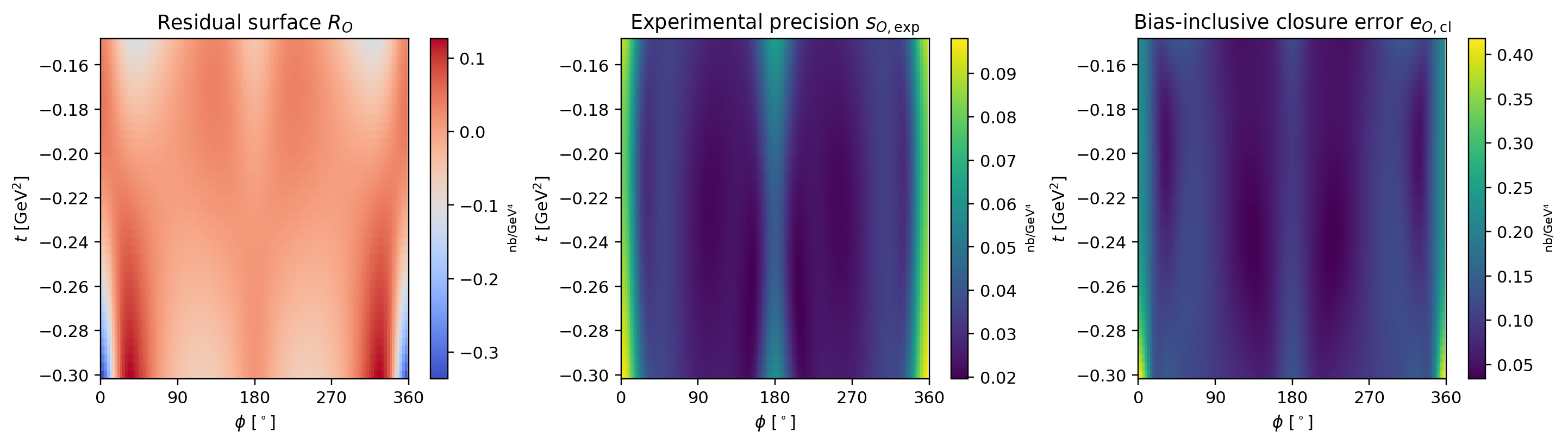}
\caption{Surrogate closure-test diagnostics for the unpolarized cross-section surface, plotted versus $t$ and the physical azimuthal angle $\phi_{\mathrm{deg}}\in[0,360^\circ]$.  The left panel is the residual $R_O$, the middle panel is the pointwise experimental precision $s_{O,\expd}$ of the nested replica means, and the right panel is the bias-inclusive closure error $e_{O,\mathrm{cl}}$ of Eq.~\eqref{eq:surrogate_total}.  All three cross-section quantities are in nb/GeV$^4$.}
\label{fig:surrogate_error_surfaces}
\end{figure*}

The surface-level closure study summarized in Fig.~\ref{fig:surrogate_error_surfaces} and Table~\ref{tab:surrogate_reporting} evaluates the pseudodata truth model at $E_{\mathrm{beam}}=5.75~\mathrm{GeV}$, $Q^2=1.50~\mathrm{GeV}^2$, and $x_B=0.22$.  The sparse design contains six $t$ slices in $[-0.30,-0.15]~\mathrm{GeV}^2$ and 24 unique azimuthal points in $[0,360^\circ)$.  The nested ensemble uses 40 experimental replicas and three stochastic surrogate trainings per replica.  Methodological spread is evaluated over eight variants specified in advance: the nominal, narrower, wider, shallower, and deeper networks, a ReLU variant, and lower- and higher-$L_2$ variants---and coverage is evaluated with 20 independent outer closure trials.  The observed training-failure rate was zero across the 896 fits.

\begin{table*}[t]
\centering
\caption{Observable-surrogate closure diagnostic from the surface-level pseudodata study.  Cross-section MAE, RMSE, and uncertainty components are in nb/GeV$^4$; the BSA quantities are dimensionless.  Coverage is the empirical pointwise coverage of the nominal $68\%$ variance interval over 20 independent outer closure trials.}
\label{tab:surrogate_reporting}
\setlength{\tabcolsep}{3.2pt}
\begin{tabular}{lccccccc}
\toprule
Observable & validation grid & $\mathrm{MAE}$ & $\mathrm{RMSE}$ & $\langle s_{O,\expd}\rangle$ & $\langle s_{O,\alg}\rangle$ & $\langle s_{O,\meth}\rangle$ & coverage \\
\midrule
$\sigma_{UU}$ & $41\times120$ $(t,\phi)$ & 0.0296 & 0.0438 & 0.0342 & 0.0316 & 0.0716 & 0.856 \\
$A_{LU}$      & $41\times120$ $(t,\phi)$ & 0.00690 & 0.00819 & 0.0122 & 0.00994 & 0.00992 & 0.864 \\
\bottomrule
\end{tabular}
\end{table*}

Figure~\ref{fig:surrogate_error_surfaces} visualizes the cross-section residual, experimental precision, and closure-error surfaces defined above, while Table~\ref{tab:surrogate_reporting} reports the corresponding domain summaries for the cross section and BSA.  The pooled pointwise coverages, 0.856 and 0.864, exceed the nominal 0.68 level.  Bootstrap resampling of the 20 independent outer-trial coverage fractions gives $95\%$ intervals $[0.828,0.881]$ and $[0.799,0.920]$, respectively.  The resulting variance assignment is therefore conservative and not yet calibrated to exact nominal coverage.  The quoted methodological term is the spread within the eight-member model family listed above, not an exhaustive theory or analysis uncertainty.

\subsection{Using surrogate surfaces in CFF extraction}

A surrogate surface ensemble can be passed to a CFF extraction in two related ways.  The governing bookkeeping rule is that every uncertainty source enters the inference chain exactly once: through an ensemble index, through a conditional covariance in the loss, or through a post-fit variation covariance.  The preferred route is to fit each surface replica directly, because the replica distribution retains non-Gaussian and nonlinear information.  A covariance-kernel fit to the mean surface on a sampled grid is a useful second-moment compression and is equivalent only to the extent that this approximation is adequate; it must retain the full correlations of Eq.~\eqref{eq:surrogate_kernel}.  For a local extraction at a kinematic point $x_0$, the direct-replica loss is
\begin{equation}
{
\begin{aligned}
  L_{\surr}^{(r)}(\theta;x_0)&=\bm d_r^T W_r^{-1}\bm d_r,\\
  d_{r,i}&=O_{\mathrm{BKM}}(\theta;x_0,\phi_i)-\widetilde O_\psi^{(r)}(x_0,\phi_i).
\end{aligned}}
  \label{eq:surrogate_local_loss}
\end{equation}
{For the preferred direct-replica route, $\widetilde O_\psi^{(r)}$ is the algorithmically averaged surface replica and $W_r$ contains only conditional observable-space uncertainty not already represented by the experimental-replica index or another explicit ensemble index.  Experimental, detector, binning, nuisance, or surrogate-training fluctuations that were sampled when constructing replica $r$ are not added again inside the same fit.  Methodological choices are propagated as end-to-end variants by default.  For the alternative mean-surface/kernel compression, a conservative effective covariance on the sampled grid is}
\begin{equation}
  {\mathbf C_{\mathrm{eff,kernel}}^{\mathrm{data}}=\mathbf K^{O,\expd}+\mathbf C_{\surr,\alg}^{\mathrm{data}}+\mathbf C_{\mathrm{bin}}^{\mathrm{data}}+\mathbf C_{\mathrm{nuis}}^{\mathrm{data}},}
  \label{eq:ceff}
\end{equation}
where {$\mathbf K^{O,\expd}$ is the discretized experimental surface kernel of Eq.~\eqref{eq:surrogate_kernel}, $\mathbf C_{\mathrm{bin}}^{\mathrm{data}}$ accounts for re-binning or bin migration not already sampled, and $\mathbf C_{\mathrm{nuis}}^{\mathrm{data}}$ contains unsampled nuisance contributions.  A methodological surface covariance may be used instead of explicit end-to-end methodological variants as a compressed second-moment alternative, but the same variation is never included both in the input covariance and again in the output methodological covariance.}  The final CFF uncertainty table should then contain both the surrogate-stage terms and the CFF-inference-stage terms.

\section{Kinematic multivariate inference}\label{sec:kmi}

The traditional local-to-global workflow fits CFFs independently at many fixed kinematic points and then trains a second model to interpolate those extracted CFFs.  This can work, but it treats local fit outputs as intermediate data and can propagate artifacts from unstable bins.  {KMI instead promotes the same four-component vector $\theta(x)$ defined in Eq.~\eqref{eq:thetavec} to one network-valued function $\theta_\Theta(x)$ trained over all kinematics.}
{For measured point $i$, the model prediction is}
\begin{equation}
  {y_i^{\mathrm{model}}=O_{\mathrm{BKM}}\left(\theta_\Theta(x_i),x_i,\phi_i\right),}
\end{equation}
with global loss
\begin{equation}
{
\begin{aligned}
  L_{\mathrm{KMI}}(\Theta)&=\bm d^T(\mathbf C^{\mathrm{data}})^{-1}\bm d+\lambda R[\theta_\Theta],\\
  d_i&=y_i^{\mathrm{model}}-y_i^{\mathrm{data}}.
\end{aligned}}
  \label{eq:kmiloss}
\end{equation}
{The common normalization factor $1/N_{\mathrm{obs}}$ has been removed so that Eqs.~\eqref{eq:chi2}, \eqref{eq:surrogate_local_loss}, and \eqref{eq:kmiloss} use the same chi-square convention; a reduced chi-square may still be reported as a diagnostic.}  {Here $\mathbf C^{\mathrm{data}}$ may include correlations across angles, kinematic bins, and observables.}  The regularizer $R$ can enforce smoothness or stability, but it must be tested with closure studies and included in the methodological uncertainty.  Model-complexity choices can be guided by validation losses and by information criteria such as AIC, BIC, or MDL.  AIC, BIC, and MDL all penalize fit quality by model complexity, but with different statistical interpretations and different penalties; therefore, in KMI they are used as comparative diagnostics over a pre-defined architecture family rather than as absolute arbiters of correctness \cite{Akaike1974,Schwarz1978,Rissanen1978,Bishop2006,Hastie2009}.  The selected criterion and the spread of near-optimal candidates are included in the methodological-error study.  KMI can be trained directly on data replicas or on surrogate surface replicas.  In surrogate-assisted KMI, each surface replica defines
\begin{equation}
{
\begin{aligned}
  L_{\mathrm{KMI}}^{(r)}(\Theta)&=\bm d_r^T W_r^{-1}\bm d_r+\lambda R[\theta_\Theta],\\
  d_{r,i}&=O_{\mathrm{BKM}}(\theta_\Theta(x_i),x_i,\phi_i)-\widetilde O_\psi^{(r)}(x_i,\phi_i).
\end{aligned}}
  \label{eq:surrogate_kmi_loss}
\end{equation}
{The output is a distribution of CFF surfaces; when repeated trainings are used, $\widetilde{\CFF}^{(r)}_a(x)$ denotes the within-replica algorithmic mean.  For the CFF-component pair $a,b$ one reports the pointwise mean and full covariance kernel,}
\begin{equation}
{
\begin{aligned}
 \delta\CFF_a^{(r)}(x)={}&
 \widetilde\CFF_a^{(r)}(x)-\overline\CFF_a(x),\\
 K^{\mathrm{CFF}}_{ab}(x,x')={}&
 \frac{1}{N_{\mathrm{rep}}-1}\sum_r
 \delta\CFF_a^{(r)}(x)\delta\CFF_b^{(r)}(x').
\end{aligned}}
\end{equation}
{Together with this kernel, one reports the closure-test bias $b_a(x)$, algorithmic surface $s_{a,\alg}(x)$, methodological surface $s_{a,\meth}(x)$, and coverage.}  In a real analysis the truth is not known, so the closure-tested bias and methodological spread from realistic pseudodata families define the assigned systematic component.

The KMI uncertainty budget contains several terms that are absent or hidden in a purely local analysis.  The shared kinematic model can reduce variance by borrowing information across neighboring bins, but it can also introduce smoothness bias if the architecture or regularizer is too restrictive.  Closure tests should therefore include truth surfaces with different smoothness scales, localized structures, and correlations among CFFs.  The following scans should be treated as methodological variations: network depth and width, activation choice, regularization strength, input normalization, observable subset, covariance model, surrogate-grid density, treatment of normalization nuisance parameters, and the kinematic domain retained for training.  For each variation, one stores the CFF residual surface and the observable residual surface.  The final KMI table should include pointwise errors and domain summaries, for example the mean absolute bias, maximum absolute bias, mean standard deviation, and coverage over the supported kinematic domain.

{When KMI is trained directly on experimental data replicas, or directly on surrogate surface replicas, the experimental fluctuation is carried by the replica index and is not inserted again into the within-replica covariance.  The matrix $W_r$ contains only conditional interpolation, detector, binning, or nuisance contributions not already sampled in replica $r$.  If the ensemble is instead compressed to a mean surface and covariance kernel, the experimental kernel enters the effective covariance as in Eq.~\eqref{eq:ceff}.  Methodological variants are propagated end to end unless they are deliberately replaced by a compressed methodological kernel; the two representations are never combined for the same source.  In either route the full grid correlations are retained rather than replaced by independent diagonal errors.}

The CFF-and-DVCS-term surface closure study uses only the unpolarized cross-section surrogate at $E_{\mathrm{beam}}=5.75~\mathrm{GeV}$ and $Q^2=1.50~\mathrm{GeV}^2$.  The input support contains six $x_B$ points in $[0.18,0.35]$, six $t$ points in $[-0.30,-0.15]~\mathrm{GeV}^2$, and 24 unique azimuthal points.  CFF outputs are summarized on a fully supported $41\times41$ $(x_B,t)$ grid.  The ensemble uses 20 experimental replicas, two surrogate-training repeats per replica, two CFF-inference repeats per surrogate, eight architecture and regularization variants specified in advance, and five independent outer closure trials.  No failed optimizations were observed among the 166 fits.

The weighted observable Jacobian has singular values $330.05$, $4.169$, $0.249$, and $0.102$.  Its effective rank is four, but its condition number is approximately $3.24\times10^3$.  The reduced four-output inverse problem is therefore formally identifiable yet numerically poorly conditioned.  No truth-tuned prior was imposed.  The large uncertainty components and conservative coverage in Table~\ref{tab:kmi_reporting} should consequently be read as an identifiability diagnostic, not as phenomenological CFF constraints.

\begin{table*}[t]
\centering
\caption{Closure diagnostic for the three real CFF surfaces and the effective azimuthally independent DVCS cross-section term, using only the unpolarized cross-section surrogate.  Entries are domain means, with maxima in parentheses, over the supported fixed-$Q^2$ and fixed-beam-energy pseudodata domain.  The $\ReH$, $\ReE$, and $\ReHt$ rows are dimensionless; the $\sigDVCS$ row is in nb/GeV$^4$.  Coverage is pooled pointwise coverage over five independent outer closure trials.  The large uncertainties reflect the poor conditioning of the reduced four-output inverse problem and are not phenomenological CFF constraints.}
\label{tab:kmi_reporting}
\scriptsize
\setlength{\tabcolsep}{2.0pt}
\begin{tabular}{lcccccc}
\toprule
Fitted output
& \shortstack{$\langle s_{\expd}\rangle$\\(max)}
& \shortstack{$\langle s_{\surr,\alg}\rangle$\\(max)}
& \shortstack{$\langle s_{\CFF,\alg}\rangle$\\(max)}
& \shortstack{$\langle s_{\meth}\rangle$\\(max)}
& \shortstack{$\langle |b|\rangle$\\(max)}
& coverage \\
\midrule
$\ReH(x)$      & 1.86 (3.09) & 2.48 (4.04) & 2.54 (3.84) & 4.56 (7.56) & 2.26 (5.32) & 0.92 \\
$\ReE(x)$      & 2.23 (2.42) & 3.29 (3.84) & 2.83 (3.56) & 3.37 (3.97) & 3.97 (4.59) & 0.80 \\
$\ReHt(x)$     & 1.58 (3.04) & 2.71 (4.45) & 2.31 (3.66) & 2.57 (4.25) & 1.24 (2.29) & 0.78 \\
$\sigDVCS(x)$  & 0.0267 (0.118) & 0.0409 (0.156) & 0.0270 (0.133) & 0.0630 (0.283) & 0.0367 (0.248) & 0.77 \\
\bottomrule
\end{tabular}
\end{table*}

The methodological column is the spread within the tested eight-member network family and is not an exhaustive uncertainty over all possible generators, BKM reductions, covariance models, or support choices.  The displayed coverage fractions are pooled over five outer trials and 1681 correlated grid points per trial.  Because the grid-point indicators are correlated and only five outer trials are available, these values are reported as descriptive pooled coverage rather than as independently calibrated binomial estimates.  All four values lie above the nominal 0.68 level, indicating nonuniform overcoverage together with weak numerical identifiability.  Split-half checks also vary appreciably for several real-CFF components, so the entries are rounded conservatively and are interpreted as closure and identifiability diagnostics rather than precision uncertainty determinations.  A quantitative extraction requires complementary observables, explicit identifiability tests, and larger converged ensembles.

\section{Event-level data, recursive re-binning, and information retention}\label{sec:event}

The local closure-test example is limited by published binned data.  If event-level data or sufficiently detailed compressed replicas are available, the extraction can use detector-level covariance and recursive re-binning.  For a Hall A-like event-level study, a detector variable vector can be written schematically as
\begin{equation}
  u=(p,\theta_e,\varphi_e,\theta_\gamma,\varphi_\gamma),
\end{equation}
where $p,\theta_e,\varphi_e$ are reconstructed electron quantities and $\theta_\gamma,\varphi_\gamma$ are photon angles.  A detector-level covariance matrix has the form
\begin{equation}
V=
\begin{pmatrix}
\sigma_p^2 & \sigma_{p\theta_e} & \sigma_{p\varphi_e} & \sigma_{p\theta_\gamma} & \sigma_{p\varphi_\gamma} \\
\sigma_{p\theta_e} & \sigma_{\theta_e}^2 & \sigma_{\theta_e\varphi_e} & \sigma_{\theta_e\theta_\gamma} & \sigma_{\theta_e\varphi_\gamma} \\
\sigma_{p\varphi_e} & \sigma_{\theta_e\varphi_e} & \sigma_{\varphi_e}^2 & \sigma_{\varphi_e\theta_\gamma} & \sigma_{\varphi_e\varphi_\gamma} \\
\sigma_{p\theta_\gamma} & \sigma_{\theta_e\theta_\gamma} & \sigma_{\varphi_e\theta_\gamma} & \sigma_{\theta_\gamma}^2 & \sigma_{\theta_\gamma\varphi_\gamma} \\
\sigma_{p\varphi_\gamma} & \sigma_{\theta_e\varphi_\gamma} & \sigma_{\varphi_e\varphi_\gamma} & \sigma_{\theta_\gamma\varphi_\gamma} & \sigma_{\varphi_\gamma}^2
\end{pmatrix}.
\label{eq:detcov}
\end{equation}
The event-level illustration uses Hall A-like scales: {small relative spectrometer momentum uncertainty}, milliradian electron angular resolution, and coarser photon angular resolution controlled by calorimeter granularity \cite{Defurne2015}.  {In a real analysis the entries of $V$ should be supplied by the experimental collaboration from its calibrated optics, reconstruction, and systematic studies; an external analyst is not being asked to reconstruct the covariance from detector-design documents.}

Derived variables $f\in\{\xb,Q^2,t,\phi\}$ are functions of $u$.  To first order,
\begin{equation}
  \sigma_f^2=\sum_{i,j}\frac{\partial f}{\partial u_i}V_{ij}\frac{\partial f}{\partial u_j}.
\end{equation}
For example, $Q^2=4Ep\sin^2(\theta_e/2)$ gives
\begin{equation}
  \sigma_{Q^2}^2=\left[4E\sin^2(\theta_e/2)\right]^2\sigma_p^2
  +\left[2Ep\sin\theta_e\right]^2\sigma_{\theta_e}^2+\cdots,
\end{equation}
where the omitted terms are covariance contributions.  This resolution sets a lower bound on meaningful bin widths and indicates where bin migration must be propagated.

A complete event-level covariance model should also contain nuisance directions that are not represented by simple detector resolutions.  Examples include luminosity normalization, acceptance corrections, radiative corrections, accidental and physics backgrounds, exclusivity-cut efficiency, trigger efficiency, beam-energy uncertainty, form-factor uncertainties used in the BH term, and false-asymmetry contributions for spin observables.  Some of these effects are global normalization shifts, some are correlated shape distortions in $\phi$, and some are local event-migration effects in $(x_B,Q^2,t,\phi)$.  In practice they can be encoded either as nuisance parameters in the loss or as additional covariance terms of the form {$s_i^{[\lambda]}s_j^{[\lambda]}$} in Eq.~\eqref{eq:covmodel}.  Event-level access is valuable because these sources can be propagated before the data are compressed into publication bins.

Recursive re-binning uses the same events with controlled changes in bin widths and starting points.  The pseudodata study uses $N_{\mathrm{evt}}=3000$ generated events per fixed kinematic chunk.  One can vary $t$ and $\xb$ bin widths while keeping the mean kinematics nearly fixed, or shift the starting point of the $\phi$ bins to obtain non-overlapping or partially shifted angular partitions.  Adaptive and Bayesian binning ideas provide useful general context for this type of recursive resampling \cite{Scargle2013,Worley2015,SanftOthmer2015}.  Each re-binning configuration is paired with covariance replicas and re-fitted.  The resulting spread becomes part of $C_{\mathrm{bin}}$ in Eq.~\eqref{eq:ceff}.

\begin{figure}[t]
\centering
\includegraphics[width=\linewidth]{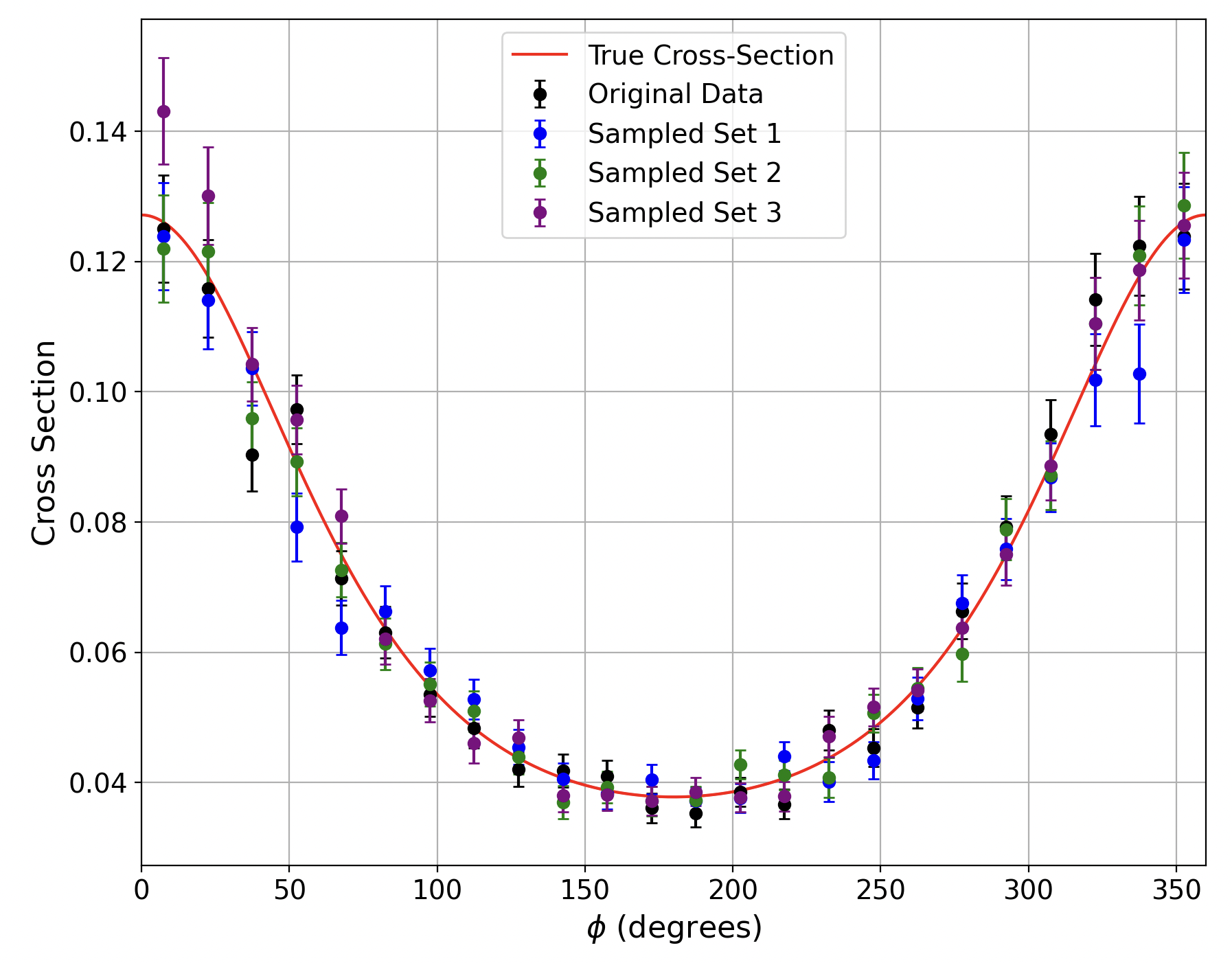}
\caption{Event-level resampling example.  The sampling iterations have the same mean kinematics but slightly different bin widths in $t$ and $x_B$.}
\label{fig:samebins}
\end{figure}

\begin{figure}[t]
\centering
\includegraphics[width=\linewidth]{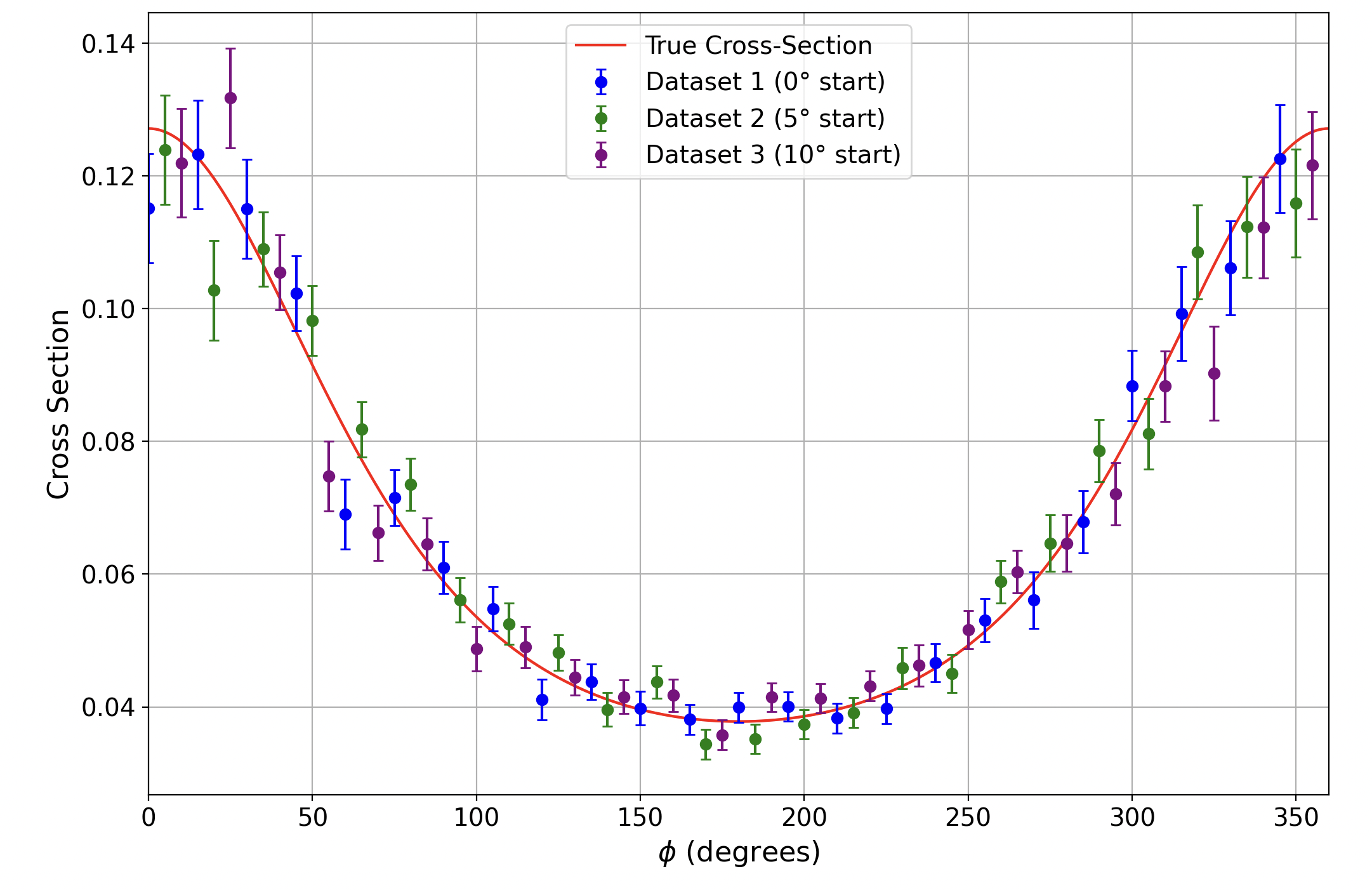}
\caption{Angular re-binning example.  Shifting the starting point of the $\phi$ binning exposes bin-migration sensitivity and creates alternative statistically valid partitions of the same event sample.}
\label{fig:phishift}
\end{figure}

\begin{figure}[t]
\centering
\includegraphics[width=\linewidth]{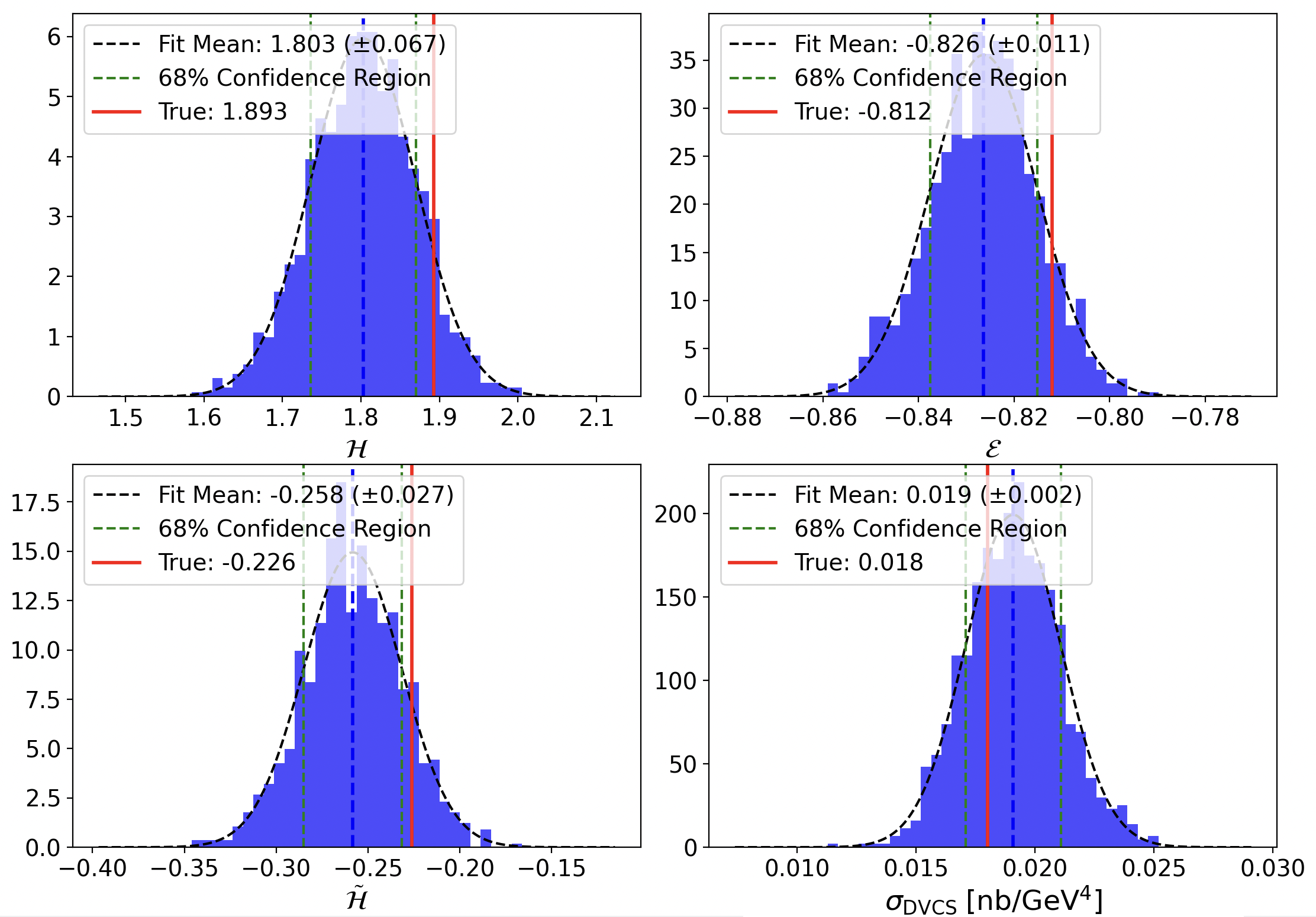}
\caption{CFF extraction with event-level resampling and recursive re-binning.  The three real-CFF panels are dimensionless, while the $\sigDVCS$ panel is in nb/GeV$^4$.  The model architecture is unchanged relative to the local-fit study; the improvement arises from better use of the available experimental information.}
\label{fig:resampled}
\end{figure}

{Figures~\ref{fig:samebins} and~\ref{fig:phishift} illustrate width and boundary variations of the same event sample, while Fig.~\ref{fig:resampled} shows the resulting change in the pseudodata CFF extraction.}

The information-theoretic motivation is straightforward.  Binning discards within-bin event locations and therefore reduces Fisher information relative to an unbinned or re-binnable representation \cite{Cramer1946,Rao1945,LehmannCasella1998}.  The Fisher-information discussion motivates the use of unbinned data, covariance replicas, and re-binnable summaries as the natural data products for statistically efficient inference.

\section{Practical uncertainty-propagation workflow}\label{sec:workflow}

{The practical workflow below links each analysis step to its statistical purpose.  Reproducing the input covariance preserves correlations; nested ensembles separate experimental and training variability; pseudodata expose bias; controlled variations quantify methodological sensitivity; and coverage tests assess whether the quoted intervals have their stated interpretation.  The implementation details differ for local fits, KMI, surrogate-assisted extraction, and event-level analysis.}

\subsection{Local-fit workflow}

\begin{enumerate}[leftmargin=*]
\item Define the observable vector and covariance matrix for the fixed kinematic point.  State whether the covariance is diagonal, block diagonal, or fully correlated, and identify missing correlations.
\item Generate BKM pseudodata with known CFFs using the same angular support and uncertainty scale as the real data.
\item For the preferred nested implementation, generate $N_{\mathrm{rep}}$ covariance replicas, train $N_{\alg}$ stochastic realizations within each replica, and use Eqs.~\eqref{eq:cffprecision} and~\eqref{eq:algerr} to obtain the experimental and algorithmic CFF covariance matrices.
\item For targeted tests, one model may be trained per covariance replica and paired with a fixed-data retraining ensemble.  Use the decomposition of Eq.~\eqref{eq:non_nested_decomp} only after the fixed-prescription, stability, and positive-semidefiniteness checks are satisfied.
\item Repeat the extraction over generator variations, BKM reductions, architecture and optimizer prescriptions, validation and stopping strategies, regularization choices, and covariance assumptions to obtain $s_{\meth}$.
\item Report the CFF mean, $s_{\expd}$, $s_{\alg}$, $s_{\meth}$, signed closure-test bias, total covariance, closure RMSE, non-Gaussian diagnostics, and CFF correlation matrix.
\end{enumerate}

\subsection{KMI workflow}

\begin{enumerate}[leftmargin=*]
\item Construct a global observable vector over all kinematic settings and assemble the covariance across angles, bins, and observables.
\item Train a CFF surface model through the BKM observable layer, using either direct data replicas or surrogate replicas.
\item Perform closure tests on multiple truth surfaces, including smooth, rapidly varying, and weakly constrained CFF components.
\item Quantify pointwise and domain-level $s_{\expd}(x)$, $s_{\alg}(x)$, $s_{\meth}(x)$, bias $b(x)$, covariance kernels, and coverage.
\item Tabulate representative benchmark kinematic points and also provide domain summaries such as maximum bias and average precision over the supported kinematic region.
\end{enumerate}

\subsection{Surrogate-assisted workflow}

\begin{enumerate}[leftmargin=*]
\item Train symmetry-constrained surrogate surfaces for the measured observables, with the appropriate even/odd $\phi$ projections applied at the output level.
\item For the preferred nested implementation, train repeated surrogate realizations within every experimental replica, average within each replica, and compute the experimental and algorithmic kernels with Eqs.~\eqref{eq:surrogate_kernel} and~\eqref{eq:surrogate_nested_kernels}.
\item For targeted tests, one surrogate may be trained per experimental replica and paired with fixed-data retraining.  Use Eq.~\eqref{eq:surrogate_non_nested} only after the compatibility and positive-semidefiniteness checks are satisfied.
\item Run closure tests on multiple BKM-generated truth surfaces and controlled methodological variants; verify that residuals are acceptable on the data-supported domain and mark unsupported regions as extrapolation.
\item Propagate either the direct surface replicas or the mean-surface/kernel compression.  In the direct route, the experimental spread remains between replicas and is not added to $W_r$; in the compressed route, the experimental kernel enters Eq.~\eqref{eq:ceff}.  Each uncertainty source is represented exactly once.
\item Report the observable-surface uncertainty table and covariance kernel before the CFF table so that data, surrogate, and CFF-inference contributions remain distinguishable.
\end{enumerate}

\subsection{Event-level and re-binnable workflow}

\begin{enumerate}[leftmargin=*]
\item Build detector-level replicas by sampling reconstructed event variables from their covariance matrix, including correlated nuisance directions where available.
\item Reconstruct $(x_B,Q^2,t,\phi)$ for each event replica and apply the same exclusivity and acceptance criteria as the experiment.
\item Generate alternative binning configurations: fixed-mean width variations in $x_B$ and $t$, shifted $\phi$ boundaries, adaptive bins, and any experiment-specific binning variants.
\item Compute cross sections or asymmetries for each replica and binning configuration, preserving correlations among bins and configurations.
\item Propagate the resulting binned or unbinned replicas to local fits, KMI, or surrogate training.  The spread across binning configurations defines the binning covariance term $C_{\rm bin}$.
\item Report detector covariance assumptions, bin-migration diagnostics, re-binning stability, and the contribution of $C_{\rm bin}$ to the final CFF budget.
\end{enumerate}

\subsection{Coverage and acceptance criteria}

For a nominal $68\%$ interval, the truth should fall inside the extracted interval in about $68\%$ of independent closure tests, within binomial uncertainty.  Coverage should be checked pointwise, for benchmark kinematics, and for integrated or simultaneous regions.  If coverage fails, the response is to identify the source: missing covariance, underpowered surrogate, unstable CFF model, BKM reduction bias, bin migration, normalization treatment, or regularization bias.  The final uncertainty assignment should then be updated by changing the model or by adding the corresponding methodological component.

A useful acceptance checklist is: (i) replica data reproduce the input covariance; (ii) algorithmic error is subdominant; (iii) surrogate residuals are smaller than or comparable to experimental precision over the supported domain; (iv) CFF closure bias is tabulated; (v) CFF correlations and covariance kernels are reported; (vi) non-Gaussian ensembles use percentile or robust intervals; (vii) bin-migration effects are propagated when event-level data are available; and (viii) methodological variations are documented in the final table.

\subsection{Obtaining CFF surfaces from surrogate observable replicas}
\label{sec:surrogate_to_cff_surface}

Once the experimental-data surrogate has been trained and validated, the next step is to propagate the surrogate ensemble into a distribution of CFF surfaces.  Let
\begin{equation}
  \widetilde{O}^{(r)}_{\psi}(u),
  \qquad
  u=(\xb,t,Q^2,\phi),
\end{equation}
be the algorithmically averaged, symmetry-constrained observable surface obtained from the $r$th covariance replica of the measured data.  In the CFF-surface study only the unpolarized cross-section surrogate $\sigma_{\uu}$ is propagated to the CFF surface, and $\theta_\Theta(x)$ uses the same four-component ordering and kinematic coordinates defined in Eq.~\eqref{eq:thetavec}.  The BSA surrogate is not an input to this four-parameter extraction.  A stacked vector such as $(\sigma_{\uu},A_{\lu},\ldots)$ is supported by the general framework only after the CFF vector and helicity-dependent BKM layer are expanded to the relevant real and imaginary components and the joint observable covariance is supplied.  The observable prediction at any point $(x,\phi)$ is computed through the corresponding deterministic BKM layer.  For each surrogate replica, the fitted CFF surface is obtained by minimizing
\begin{equation}
{
\begin{aligned}
 L^{(r)}_{\surr\rightarrow\CFF}(\Theta)={}&
 \sum_{i,j}d^{(r)}_i
 [\mathbf C^{(r),\mathrm{data}}_{\mathrm{eff}}]^{-1}_{ij}
 d^{(r)}_j\\
 &+\lambda R[\theta_{\Theta}],\\
 d^{(r)}_i={}&O^{\mathrm{BKM}}_i
 (\theta_{\Theta}(x_i),x_i,\phi_i)\\
 &-\widetilde O^{(r)}_{\psi}(x_i,\phi_i).
\end{aligned}}
\label{eq:surrogate_to_cff_loss}
\end{equation}
{Here the indices $i,j$ run over all sampled kinematic, angular, and observable entries used in the surface fit.  For the direct-replica route, $\mathbf C^{(r),\mathrm{data}}_{\mathrm{eff}}$ contains only conditional surrogate interpolation covariance and non-replicated binning, detector, or nuisance covariance.  The experimental spread carried between surrogate replicas is not added again within the same fit.  If surrogate-training repeats are propagated explicitly as an additional ensemble index rather than averaged, their covariance is likewise omitted from $\mathbf C^{(r),\mathrm{data}}_{\mathrm{eff}}$.  For the alternative mean-surface/kernel compression, the experimental surface kernel enters through Eq.~\eqref{eq:ceff}.}
\begin{equation}
  {\mathbf C^{(r),\mathrm{data}}_{\mathrm{eff}}
  =\mathbf C_{\surr,\alg}^{(r),\mathrm{data}}+\mathbf C_{\mathrm{bin}}^{\mathrm{data}}+\mathbf C_{\mathrm{nuis}}^{\mathrm{data}} .}
  \label{eq:cff_surface_effective_covariance}
\end{equation}
The term {$\mathbf C_{\mathrm{nuis}}^{\mathrm{data}}$} represents correlated nuisance contributions such as normalization, beam polarization, acceptance, or calibration effects when they are available as nuisance parameters rather than fixed errors and have not already been sampled in the replica.  The regularizer $R[\theta_{\Theta}]$ may enforce smoothness or stability of the CFF surface, but its strength and form are treated as end-to-end methodological variations and are not also inserted into the conditional covariance.

{Training Eq.~\eqref{eq:surrogate_to_cff_loss} for all $N_{\mathrm{rep}}$ surrogate replicas and repeated CFF-inference trainings $q=1,\ldots,N_{\alg}$ gives surfaces $\CFF^{(r,q)}_a(x)$.  The experimental-replica surface is the algorithmic mean within replica $r$,}
\begin{equation}
{
\begin{aligned}
 \widetilde{\CFF}^{(r)}_a(x)={}&
 \frac{1}{N_{\alg}}\sum_{q=1}^{N_{\alg}}\CFF^{(r,q)}_a(x),\\
 \overline{\CFF}_a(x)={}&
 \frac{1}{N_{\mathrm{rep}}}\sum_{r=1}^{N_{\mathrm{rep}}}
 \widetilde{\CFF}^{(r)}_a(x).
\end{aligned}}
  \label{eq:cff_surface_mean}
\end{equation}
{Here $a\in\{\ReH,\ReE,\ReHt,\sigDVCS\}$.  Because a CFF surface is a correlated object, the experimental-replica covariance kernel is computed from those within-replica means,}
\begin{equation}
{
\begin{aligned}
  K^{\mathrm{CFF},\expd}_{ab}(x,x')={}&
  \frac{1}{N_{\mathrm{rep}}-1}
  \sum_{r=1}^{N_{\mathrm{rep}}}
  \delta_r\CFF_a(x)\,\delta_r\CFF_b(x'),\\
  \delta_r\CFF_a(x)={}&
  \widetilde{\CFF}^{(r)}_a(x)-\overline{\CFF}_a(x),\\
  s_{\expd,a}(x)={}&\sqrt{K^{\mathrm{CFF},\expd}_{aa}(x,x)}.
\end{aligned}}
  \label{eq:cff_surface_kernel}
\end{equation}
The diagonal of this kernel gives the plotted pointwise {standard deviation}, while the off-diagonal structure is required for any later integration, interpolation, comparison with theory, or propagation to GPD-level quantities.

The CFF-surface extraction has two algorithmic components.  The first is the surrogate interpolation variability, already quantified during surrogate training.  The second is the CFF-inference variability obtained by retraining Eq.~\eqref{eq:surrogate_to_cff_loss} on the same surrogate surface with independent initialization, optimization history, and batching.  If $q=1,\ldots,N_{\alg}$ labels repeated trainings on fixed surrogate replicas, then
\begin{equation}
{
\begin{aligned}
 K^{\mathrm{CFF},\alg}_{ab}(x,x')={}&
 \frac{1}{N_{\mathrm{rep}}}\sum_{r=1}^{N_{\mathrm{rep}}}
 \frac{1}{N_{\alg}-1}\\
 &\times\sum_{q=1}^{N_{\alg}}
 \delta_q\CFF^{(r,q)}_a(x)
 \delta_q\CFF^{(r,q)}_b(x'),\\
 \delta_q\CFF^{(r,q)}_a(x)={}&
 \CFF^{(r,q)}_a(x)-\widetilde\CFF^{(r)}_a(x),\\
 s_{\CFF,\alg,a}(x)={}&
 \sqrt{K^{\mathrm{CFF},\alg}_{aa}(x,x)}.
\end{aligned}}
  \label{eq:cff_surface_alg}
\end{equation}
Methodological uncertainty is evaluated by repeating the full closure-test chain under controlled variations of the generator family, BKM reduction, regularization, architecture, surrogate grid, observable subset, support mask, and covariance model.  For a known pseudodata truth surface $\CFF^{\truth}_a(x)$, the closure bias is
\begin{equation}
  b_a(x)=\overline{\CFF}_a(x)-\CFF^{\truth}_a(x),
  \label{eq:cff_surface_bias}
\end{equation}
{while the methodological covariance kernel is computed directly from the distribution of signed residual surfaces over the accepted variations,}
\begin{equation}
{
\begin{aligned}
 K^{\mathrm{CFF},\meth}_{ab}(x,x')={}&
 \frac{1}{N_m-1}\sum_{m=1}^{N_m}
 \delta b^{(m)}_a(x)\delta b^{(m)}_b(x'),\\
 \delta b^{(m)}_a(x)={}&
 b^{(m)}_a(x)-\overline b_a(x),\\
 s_{\meth,a}(x)={}&
 \sqrt{K^{\mathrm{CFF},\meth}_{aa}(x,x)}.
\end{aligned}}
  \label{eq:cff_surface_method}
\end{equation}
For real data, where $\CFF^{\truth}_a(x)$ is unknown, the closure-tested bias envelope is carried as a methodological component and reported together with the experimental and algorithmic terms.

The kernel $K^{\mathrm{CFF},\surr,\alg}_{ab}$ used below denotes the CFF-space covariance induced by the surrogate interpolation component, propagated either through the conditional covariance or through an explicit surrogate-training ensemble.  It is omitted as a separate addition when that same variability is already contained in another propagated ensemble component.

{Rather than repeat the scalar quadrature of Eq.~\eqref{eq:total_nobias}, the surface uncertainty is assembled at kernel level,}
\begin{equation}
{
\begin{aligned}
 K^{\mathrm{CFF},\tot}_{ab}={}&
 K^{\mathrm{CFF},\expd}_{ab}
 +K^{\mathrm{CFF},\surr,\alg}_{ab}\\
 &+K^{\mathrm{CFF},\alg}_{ab}
 +K^{\mathrm{CFF},\meth}_{ab}\\
 &+K^{\mathrm{CFF},\cov}_{ab}
 +K^{\mathrm{CFF},\mathrm{supp}}_{ab}\\
 &+K^{\mathrm{CFF},\mathrm{cross}}_{ab},\\
 s_{\tot,a}(x)={}&
 \sqrt{K^{\mathrm{CFF},\tot}_{aa}(x,x)}.
\end{aligned}}
  \label{eq:cff_surface_total_uncertainty}
\end{equation}
{Here the covariance-model and support kernels are obtained from their corresponding variation ensembles, and the cross-kernel is retained whenever the components are not empirically independent.}  The signed bias $b_a(x)$ should be quoted separately.  For closure studies it is useful to also report the root-mean-square closure error,
\begin{equation}
  e_{\mathrm{cl},a}(x)=
  \sqrt{b_a^2(x)+s^2_{\tot,a}(x)} .
  \label{eq:cff_surface_closure_error}
\end{equation}

\begin{figure}[htbp]
\centering
\includegraphics[width=\columnwidth]{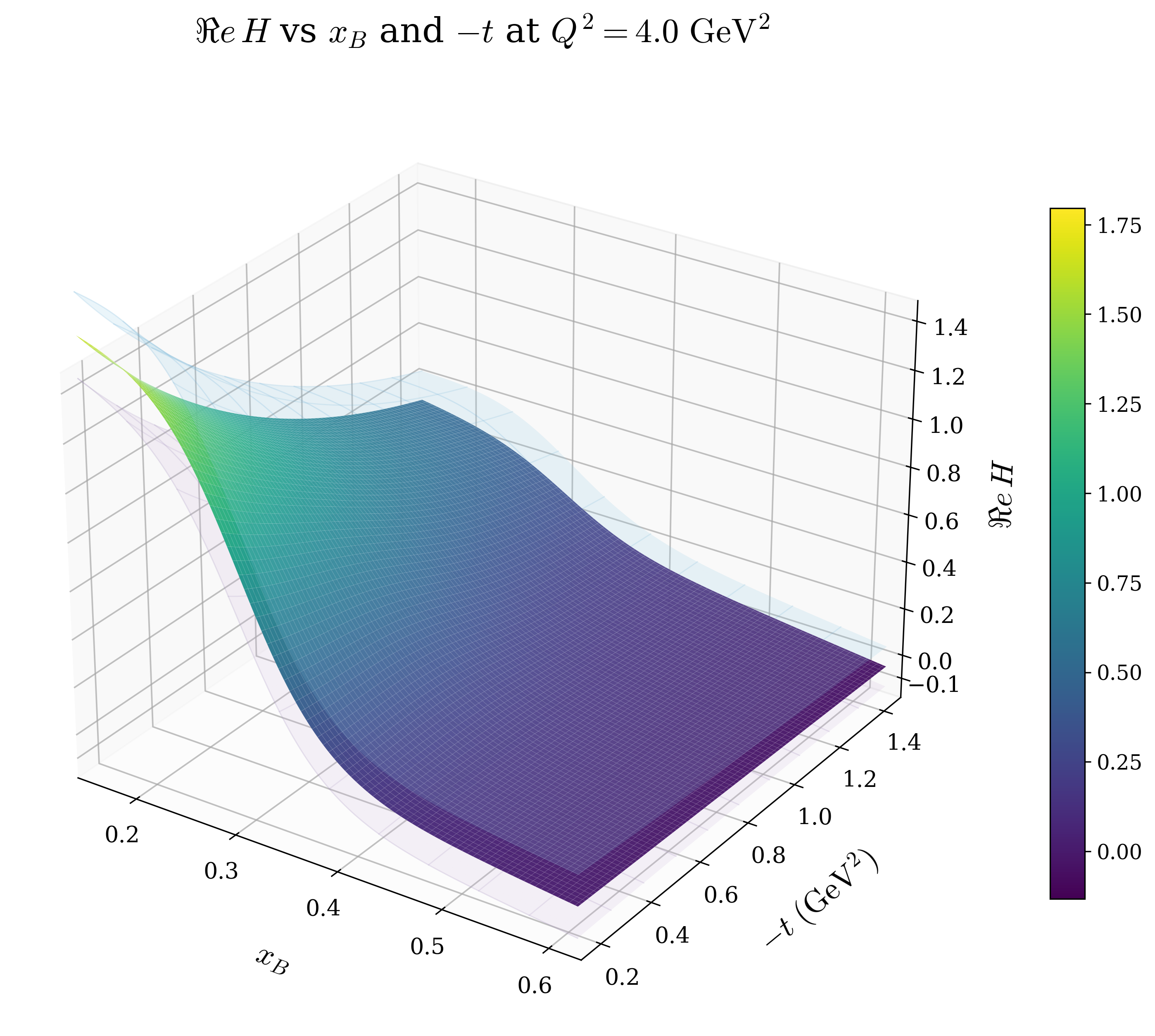}
\caption{Example CFF surface obtained from a surrogate-assisted extraction.
The central surface shows the mean $\ReH(\xb,-t)$ at fixed
$Q^2=4.0~\mathrm{GeV}^2$. The translucent upper and lower sheets show the
pointwise $\pm1\sigma$ uncertainty envelope. In a full analysis, this envelope
is computed from the experimental replica spread, surrogate interpolation
error, CFF-inference algorithmic error, methodological variation,
covariance-model uncertainty, and support penalty, as in
Eq.~\eqref{eq:cff_surface_total_uncertainty}.}
\label{fig:reh_surface_from_surrogate}
\end{figure}

Figure~\ref{fig:reh_surface_from_surrogate} illustrates the final form of this result for $\ReH$ at fixed $Q^2=4.0~\mathrm{GeV}^2$.  The central surface represents $\overline{\ReH}(\xb,-t;Q^2)$, while the translucent upper and lower sheets represent the pointwise $\pm1\sigma$ envelope from Eq.~\eqref{eq:cff_surface_total_uncertainty}.  The plotted envelope is a compact visualization of the local uncertainty scale; the covariance kernel in Eq.~\eqref{eq:cff_surface_kernel} is the object that should be used for quantitative propagation.

For reporting, this surface-level result should be tabulated at benchmark kinematic points and summarized over the supported domain.  A minimal table should include $\overline{\CFF}_a$, $s_{\expd,a}$, $s_{\surr,\alg,a}$, $s_{\CFF,\alg,a}$, $s_{\meth,a}$, $s_{\mathrm{cov},a}$, $s_{\mathrm{supp},a}$, the closure bias $b_a$ when available, and the coverage fraction from independent pseudodata tests.  This table should be supplied together with the covariance kernel or a compressed representation of it, since the visual $1\sigma$ envelope alone does not encode the correlations across the CFF surface.

\section{Data products needed from experiments}\label{sec:datarelease}

{Because data-release constraints differ among experiments, the recommended products are organized into three cumulative tiers:}
\begin{enumerate}[leftmargin=*]
\item {\textit{Tier 1 (minimum):} published central values, bin definitions, and the full covariance matrix across released bins and observables, together with the definitions of included systematic sources.}
\item {\textit{Tier 2:} Tier 1 plus covariance replicas or a replica-generation prescription, signed nuisance-response vectors, migration or response matrices, and metadata sufficient to reproduce the reported observable.}
\item {\textit{Tier 3 (preferred where feasible):} Tier 2 plus event-level or re-binnable data and the calibrated detector response needed to reconstruct $(\xb,Q^2,t,\phi)$ and the exclusivity selections.}
\end{enumerate}
{This tiering preserves a useful minimum even when event-level release is impractical.  Tables~\ref{tab:local_budget_expanded}, \ref{tab:method_budget_map}, \ref{tab:surrogate_reporting}, and~\ref{tab:kmi_reporting} provide examples of compact uncertainty products that can be supplied in the article or supplementary material.  The rationale is not specific to neural networks: retaining covariance prevents later analysts from replacing measured correlations with assumptions.}

The tiers also define increasing levels of interoperability for a cumulative global surrogate.  Tier~1 permits new measurements to enter a correlated world-data fit; Tier~2 supports statistically faithful replica-level updates with explicit nuisance and migration information; and Tier~3 permits re-binning and detector-level propagation as the global representation is refined.  The available tier therefore determines how completely a new experiment can update the shared surrogate without discarding the uncertainty structure or provenance of measurements already included.

\section{Summary}\label{sec:summary}

This manuscript provides a CFF-only methodology for propagating experimental and methodological uncertainty through DNN-assisted DVCS extraction.  The uncertainty budget is developed for four practical workflows: local extraction at fixed kinematics, KMI over many kinematic settings, surrogate-assisted inference with symmetry-constrained observable surfaces, and event-level or re-binnable analyses.  In each workflow the observable loss is evaluated through the BKM formalism and the measured covariance is propagated with statistically consistent replicas.  Nested replica--training ensembles directly separate experimental and algorithmic covariance, while a complementary non-nested construction pairs one fit per replica with fixed-data retraining for stability and non-Gaussianity studies.

The local closure-test material presented here is organized around the quantities that must be reported: accuracy, precision, algorithmic error, methodological error, total propagated uncertainty, non-Gaussian diagnostics, and CFF correlations.  The Hall A-like local results demonstrate why these categories are needed: some CFF components remain broad, biased, or non-Gaussian even when the reconstructed observable appears satisfactory.  The surface-level pseudodata study further finds conservative coverage and a four-output Jacobian condition number of approximately $3.24\times10^3$, exposing weak CFF identifiability rather than a physical determination.  The BSA figures test only the odd-symmetry surrogate construction and are not propagated through the displayed real-only four-parameter extraction; a joint cross-section/asymmetry analysis requires the corresponding expanded real-and-imaginary CFF basis.  The surrogate workflow extends the same logic from points to surfaces by defining residual, precision, algorithmic, methodological, and total uncertainty surfaces, along with the surface covariance kernel needed for downstream CFF extraction.  KMI uses the same observable-level loss over many kinematic settings and requires pointwise and domain-level coverage tests to control smoothness bias.  Event-level and re-binnable analyses add detector covariance, bin migration, and re-binning covariance to the same budget.

Beyond individual extractions, the framework is intended to support a versioned global surrogate ensemble of DVCS observables in which new experimental releases refine the current world-data representation while retaining covariance, nuisance structure, measurement conditions, support, provenance, and unresolved inter-dataset tensions.  Each validated release would be auditable and could serve as a common observable-level input from which local or KMI CFF extractions are regenerated under a documented BKM layer.  Constructing and maintaining that global ensemble lies beyond the scope of this study, but the formalism developed here is designed to provide its statistical and data-interface foundations.

The resulting prescription is intended to be used as an uncertainty-reporting guide.  {The framework emphasizes reproducibility, covariance preservation, and validated coverage while allowing the data product to be adapted to the practical constraints of each experiment.  A complete DNN CFF result therefore consists of the central CFF values or surfaces together with the separated uncertainty components, covariance matrices or kernels, closure-test bias estimates, and coverage diagnostics.}  With these quantities in hand, neural-network CFF extraction becomes a reproducible information-extraction procedure that can be compared across experiments, architectures, and data-release formats.

\section*{Statements and Declarations}

\paragraph{Funding.}
This work was supported by the U.S. Department of Energy, Office of Science, Office of Nuclear Physics, under Award No. DE-FG02-96ER40950.

\paragraph{Data availability.}
No new experimental data were collected.  The numerical results reported in this article were obtained from generated pseudodata; the published Hall A measurement used to set the representative uncertainty scale is cited in the text.  The configurations, random seeds, tabulated numerical summaries, and supporting analysis outputs are available from the corresponding author upon reasonable request.

\paragraph{Code availability.}
The code used to generate the pseudodata, implement the BKM observable layer, train the neural-network models, propagate the uncertainty ensembles, and produce the reported tables and figures is available from the corresponding author upon reasonable request.

\paragraph{Competing interests.}
The author declares no competing interests.

\paragraph{Author contributions.}
D.K. conceived the study, developed the methodology and software, performed the analysis, interpreted the results, and wrote the manuscript.

\paragraph{Use of generative artificial intelligence.}
Overleaf AI (AI Assist, powered by Writefull) was used to assist with language editing, proofreading, typos and revision checks.

\section*{Acknowledgments}
The author acknowledges Research Computing at the University of Virginia for providing computational resources and technical support that contributed to this work (\url{https://rc.virginia.edu}).

\FloatBarrier
\IfFileExists{spphys.bst}{\bibliographystyle{spphys}}{\bibliographystyle{unsrt}}
\bibliography{reff}

@article{Mueller1994,
  author = {Mueller, D. and Robaschik, D. and Geyer, B. and Dittes, F. M. and Horejsi, J.},
  title = {Wave functions, evolution equations and evolution kernels from light-ray operators of {QCD}},
  journal = {Fortschr. Phys.},
  volume = {42},
  pages = {101--141},
  year = {1994}
}

@article{Ji1997D,
  author = {Ji, X.-D.},
  title = {Gauge-invariant decomposition of nucleon spin},
  journal = {Phys. Rev. D},
  volume = {55},
  pages = {7114--7125},
  year = {1997}
}

@article{Ji1997PRL,
  author = {Ji, X.-D.},
  title = {Deeply-virtual Compton scattering},
  journal = {Phys. Rev. Lett.},
  volume = {78},
  pages = {610--613},
  year = {1997}
}

@article{Radyushkin1996,
  author = {Radyushkin, A. V.},
  title = {Scaling limit of deeply virtual Compton scattering},
  journal = {Phys. Lett. B},
  volume = {380},
  pages = {417--425},
  year = {1996}
}

@article{Radyushkin1997,
  author = {Radyushkin, A. V.},
  title = {Nonforward parton distributions},
  journal = {Phys. Rev. D},
  volume = {56},
  pages = {5524--5557},
  year = {1997}
}

@article{BelitskyKirchnerMueller2002,
  author = {Belitsky, A. V. and Mueller, D. and Kirchner, A.},
  title = {Theory of deeply virtual Compton scattering on the nucleon},
  journal = {Nucl. Phys. B},
  volume = {629},
  pages = {323--392},
  year = {2002}
}

@article{BelitskyMueller2010,
  author = {Belitsky, A. V. and Mueller, D.},
  title = {Exclusive electroproduction revisited: treating kinematical effects},
  journal = {Phys. Rev. D},
  volume = {82},
  pages = {074010},
  year = {2010}
}

@article{BacchettaTrento2004,
  author = {Bacchetta, A. and D'Alesio, U. and Diehl, M. and Miller, C. A.},
  title = {Single-spin asymmetries: The Trento conventions},
  journal = {Phys. Rev. D},
  volume = {70},
  pages = {117504},
  year = {2004}
}

@article{Kelly2004,
  author = {Kelly, J. J.},
  title = {Simple parametrization of nucleon form factors},
  journal = {Phys. Rev. C},
  volume = {70},
  pages = {068202},
  year = {2004}
}

@article{Defurne2015,
  author = {Defurne, M. and others},
  collaboration = {Jefferson Lab Hall A},
  title = {Deeply virtual Compton scattering off the proton at 6 {GeV}},
  journal = {Phys. Rev. C},
  volume = {92},
  pages = {055202},
  year = {2015}
}

@article{Georges2022,
  author = {Georges, F. and others},
  collaboration = {Jefferson Lab Hall A},
  title = {Deeply Virtual Compton Scattering Cross Section at High Bjorken {$x_B$}},
  journal = {Phys. Rev. Lett.},
  volume = {128},
  pages = {252002},
  year = {2022}
}

@article{Kumericki2011,
  author = {Kumericki, K. and Mueller, D. and Schaefer, A.},
  title = {Neural network generated parametrizations of deeply virtual Compton form factors},
  journal = {JHEP},
  volume = {07},
  pages = {073},
  year = {2011},
  eprint = {1106.2808},
  archivePrefix = {arXiv},
  primaryClass = {hep-ph}
}

@article{Moutarde2019,
  author = {Moutarde, H. and Sznajder, P. and Wagner, J.},
  title = {Unbiased determination of {DVCS} Compton form factors},
  journal = {Eur. Phys. J. C},
  volume = {79},
  pages = {614},
  year = {2019},
  eprint = {1905.02089},
  archivePrefix = {arXiv},
  primaryClass = {hep-ph}
}

@article{Cuic2020,
  author = {Cuic, M. and Kumericki, K. and Schaefer, A.},
  title = {Separation of quark flavors using deeply virtual Compton scattering data},
  journal = {Phys. Rev. Lett.},
  volume = {125},
  pages = {232005},
  year = {2020},
  eprint = {2007.00029},
  archivePrefix = {arXiv},
  primaryClass = {hep-ph}
}

@article{CaleroDiazKeller2025,
  author = {Calero Diaz, L. and Keller, D.},
  title = {Global deep neural network modeling of Compton form factors constrained from local {$\chi^2$} maps fits},
  journal = {Phys. Rev. D},
  volume = {112},
  pages = {096001},
  year = {2025},
  doi = {10.1103/sb63-sfdt},
  eprint = {2509.18331},
  archivePrefix = {arXiv},
  primaryClass = {nucl-ex}
}

@article{LeKeller2026,
  author = {Le, Brandon B. and Keller, D.},
  title = {Compton form-factor extraction using quantum deep neural networks},
  journal = {Phys. Rev. C},
  volume = {113},
  pages = {045214},
  year = {2026},
  doi = {10.1103/p6m4-x5fg},
  eprint = {2504.15458},
  archivePrefix = {arXiv},
  primaryClass = {cs.LG}
}

@article{Fernando2023,
  author = {Fernando, I. P. and Keller, D.},
  title = {Extraction of the Sivers function with deep neural networks},
  journal = {Phys. Rev. D},
  volume = {108},
  pages = {054007},
  year = {2023},
  doi = {10.1103/PhysRevD.108.054007},
  eprint = {2304.14328},
  archivePrefix = {arXiv},
  primaryClass = {hep-ph}
}

@article{Fernando2025TMD,
  author = {Fernando, I. P. and Keller, D.},
  title = {Deep neural network extraction of unpolarized transverse momentum distributions},
  journal = {Phys. Rev. D},
  volume = {113},
  pages = {096017},
  year = {2026},
  doi = {10.1103/p3sg-k524},
  eprint = {2510.17243},
  archivePrefix = {arXiv},
  primaryClass = {hep-ph}
}

@article{BacchettaMAP2025,
  author = {Bacchetta, A. and Bertone, V. and Bissolotti, C. and Cerutti, M. and Radici, M. and Rodini, S. and Rossi, L.},
  title = {Neural-network extraction of unpolarized transverse-momentum-dependent distributions},
  journal = {Phys. Rev. Lett.},
  volume = {135},
  pages = {021904},
  year = {2025},
  doi = {10.1103/csc2-bj91},
  eprint = {2502.04166},
  archivePrefix = {arXiv},
  primaryClass = {hep-ph}
}

@article{BallNNPDF2022,
  author = {Ball, R. D. and others},
  collaboration = {NNPDF},
  title = {The path to proton structure at one-percent accuracy},
  journal = {Eur. Phys. J. C},
  volume = {82},
  pages = {428},
  year = {2022},
  eprint = {2109.02653},
  archivePrefix = {arXiv},
  primaryClass = {hep-ph}
}

@article{NNPDFClosure2014,
  author = {Deans, C. S.},
  title = {Closure testing the {NNPDF3.0} methodology},
  journal = {Nucl. Part. Phys. Proc.},
  volume = {258--259},
  pages = {25--28},
  year = {2015},
  eprint = {1409.4283},
  archivePrefix = {arXiv},
  primaryClass = {hep-ph}
}

@article{MetropolisUlam1949,
  author = {Metropolis, N. and Ulam, S.},
  title = {The Monte Carlo method},
  journal = {J. Am. Stat. Assoc.},
  volume = {44},
  pages = {335--341},
  year = {1949}
}

@article{Efron1979,
  author = {Efron, B.},
  title = {Bootstrap methods: Another look at the jackknife},
  journal = {Ann. Statist.},
  volume = {7},
  pages = {1--26},
  year = {1979}
}

@book{Press1992,
  author = {Press, W. H. and Teukolsky, S. A. and Vetterling, W. T. and Flannery, B. P.},
  title = {Numerical Recipes in {C}: The Art of Scientific Computing},
  edition = {2},
  publisher = {Cambridge University Press},
  year = {1992}
}

@book{GolubVanLoan1996,
  author = {Golub, G. H. and Van Loan, C. F.},
  title = {Matrix Computations},
  edition = {3},
  publisher = {Johns Hopkins University Press},
  year = {1996}
}

@book{Cramer1946,
  author = {Cramer, H.},
  title = {Mathematical Methods of Statistics},
  publisher = {Princeton University Press},
  year = {1946}
}

@article{Rao1945,
  author = {Rao, C. R.},
  title = {Information and the accuracy attainable in the estimation of statistical parameters},
  journal = {Bull. Calcutta Math. Soc.},
  volume = {37},
  pages = {81--91},
  year = {1945}
}

@book{LehmannCasella1998,
  author = {Lehmann, E. L. and Casella, G.},
  title = {Theory of Point Estimation},
  edition = {2},
  publisher = {Springer},
  year = {1998}
}

@article{Cybenko1989,
  author = {Cybenko, G.},
  title = {Approximation by superpositions of a sigmoidal function},
  journal = {Math. Control Signals Systems},
  volume = {2},
  pages = {303--314},
  year = {1989}
}

@article{Hornik1991,
  author = {Hornik, K.},
  title = {Approximation capabilities of multilayer feedforward networks},
  journal = {Neural Networks},
  volume = {4},
  pages = {251--257},
  year = {1991}
}

@book{Goodfellow2016,
  author = {Goodfellow, I. and Bengio, Y. and Courville, A.},
  title = {Deep Learning},
  publisher = {MIT Press},
  year = {2016}
}

@book{Bishop2006,
  author = {Bishop, C. M.},
  title = {Pattern Recognition and Machine Learning},
  publisher = {Springer},
  year = {2006}
}

@book{Hastie2009,
  author = {Hastie, T. and Tibshirani, R. and Friedman, J.},
  title = {The Elements of Statistical Learning},
  edition = {2},
  publisher = {Springer},
  year = {2009}
}

@article{Akaike1974,
  author = {Akaike, H.},
  title = {A new look at the statistical model identification},
  journal = {IEEE Trans. Autom. Control},
  volume = {19},
  pages = {716--723},
  year = {1974},
  doi = {10.1109/TAC.1974.1100705}
}

@article{Schwarz1978,
  author = {Schwarz, G.},
  title = {Estimating the dimension of a model},
  journal = {Ann. Statist.},
  volume = {6},
  pages = {461--464},
  year = {1978},
  doi = {10.1214/aos/1176344136}
}

@article{Rissanen1978,
  author = {Rissanen, J.},
  title = {Modeling by shortest data description},
  journal = {Automatica},
  volume = {14},
  pages = {465--471},
  year = {1978},
  doi = {10.1016/0005-1098(78)90005-5}
}

@article{Scargle2013,
  author = {Scargle, J. D. and Norris, J. P. and Jackson, B. and Chiang, J.},
  title = {Studies in astronomical time series analysis. {VI}. {Bayesian} block representations},
  journal = {Astrophys. J.},
  volume = {764},
  pages = {167},
  year = {2013}
}

@article{Worley2015,
  author = {Worley, B. and Powers, R.},
  title = {Generalized adaptive intelligent binning of multiway data},
  journal = {Chemom. Intell. Lab. Syst.},
  volume = {146},
  pages = {42--46},
  year = {2015}
}

@article{SanftOthmer2015,
  author = {Sanft, K. and Othmer, H. G.},
  title = {Constant-complexity stochastic simulation algorithm with optimal binning},
  journal = {J. Chem. Phys.},
  volume = {143},
  pages = {074108},
  year = {2015}
}

@book{DavisonHinkley1997,
  author = {Davison, A. C. and Hinkley, D. V.},
  title = {Bootstrap Methods and their Application},
  publisher = {Cambridge University Press},
  year = {1997},
  doi = {10.1017/CBO9780511802843}
}

@article{DelDebbio2005,
  author = {Del Debbio, L. and Forte, S. and Latorre, J. I. and Piccione, A. and Rojo, J.},
  collaboration = {NNPDF},
  title = {Unbiased determination of the proton structure function {$F_2^p$} with faithful uncertainty estimation},
  journal = {JHEP},
  volume = {03},
  pages = {080},
  year = {2005},
  doi = {10.1088/1126-6708/2005/03/080},
  eprint = {hep-ph/0501067},
  archivePrefix = {arXiv}
}

@article{DelDebbio2007,
  author = {Del Debbio, L. and Forte, S. and Latorre, J. I. and Piccione, A. and Rojo, J.},
  collaboration = {NNPDF},
  title = {Neural network determination of parton distributions: the nonsinglet case},
  journal = {JHEP},
  volume = {03},
  pages = {039},
  year = {2007},
  doi = {10.1088/1126-6708/2007/03/039},
  eprint = {hep-ph/0701127},
  archivePrefix = {arXiv}
}

@article{Ball2009,
  author = {Ball, R. D. and Del Debbio, L. and Forte, S. and Guffanti, A. and Latorre, J. I. and Piccione, A. and Rojo, J. and Ubiali, M.},
  collaboration = {NNPDF},
  title = {A determination of parton distributions with faithful uncertainty estimation},
  journal = {Nucl. Phys. B},
  volume = {809},
  pages = {1--63},
  year = {2009},
  doi = {10.1016/j.nuclphysb.2008.09.037},
  note = {Erratum: Nucl. Phys. B 816, 293 (2009)},
  eprint = {0808.1231},
  archivePrefix = {arXiv},
  primaryClass = {hep-ph}
}

@article{Geman1992,
  author = {Geman, S. and Bienenstock, E. and Doursat, R.},
  title = {Neural networks and the bias/variance dilemma},
  journal = {Neural Computation},
  volume = {4},
  number = {1},
  pages = {1--58},
  year = {1992},
  doi = {10.1162/neco.1992.4.1.1}
}

@article{Breiman1996,
  author = {Breiman, L.},
  title = {Bagging predictors},
  journal = {Machine Learning},
  volume = {24},
  number = {2},
  pages = {123--140},
  year = {1996},
  doi = {10.1007/BF00058655}
}

@inproceedings{Lakshminarayanan2017,
  author = {Lakshminarayanan, B. and Pritzel, A. and Blundell, C.},
  title = {Simple and Scalable Predictive Uncertainty Estimation using Deep Ensembles},
  booktitle = {Advances in Neural Information Processing Systems 30},
  pages = {6402--6413},
  year = {2017},
  eprint = {1612.01474},
  archivePrefix = {arXiv}
}

@book{BurnhamAnderson2002,
  author = {Burnham, K. P. and Anderson, D. R.},
  title = {Model Selection and Multimodel Inference: A Practical Information-Theoretic Approach},
  edition = {2},
  publisher = {Springer},
  address = {New York},
  year = {2002},
  doi = {10.1007/b97636}
}

@article{Cowan2019,
  author        = {Cowan, Glen},
  title         = {Statistical models with uncertain error parameters},
  journal       = {Eur. Phys. J. C},
  volume        = {79},
  pages         = {133},
  year          = {2019},
  doi           = {10.1140/epjc/s10052-019-6644-4},
  eprint        = {1809.05778},
  archivePrefix = {arXiv},
  primaryClass  = {physics.data-an}
}

\end{document}